\documentclass[fleqn,10pt]{wlscirep}
\usepackage[utf8]{inputenc}
\usepackage[T1]{fontenc}
\usepackage{comment}
\usepackage{subfigure}
\usepackage{caption}
\usepackage{multirow}
\usepackage[normalem]{ulem}
\usepackage[mathscr]{eucal}
\def\d{{\rm d}}

\title{Imprints of massive black hole binaries on neighbouring deci-Hz gravitational-wave sources}

\author[1,2,*]{Jakob Stegmann}
\author[3,4]{Lorenz Zwick}
\author[5,2]{Sander M. Vermeulen}
\author[2]{Fabio Antonini}
\author[4]{Lucio Mayer}
\affil[1]{Max Planck Institute for Astrophysics, Karl-Schwarzschild-Straße 1, 85741 Garching, Germany}
\affil[2]{Gravity Exploration Institute, School of Physics and Astronomy, Cardiff University, Cardiff, CF24 3AA, UK}
\affil[3]{Niels Bohr International Academy, Niels Bohr Institute, Blegdamsvej 17, 2100 Copenhagen, Denmark}
\affil[4]{Center for Theoretical Astrophysics and Cosmology, Institute for Computational Science, University of Zurich, Winterthurerstrasse 190, 8057 Zurich, Switzerland}
\affil[5]{California Institute of Technology, Department of Physics, Pasadena, California 91125, USA}

\affil[*]{E-mail: \href{mailto:jstegmann@mpa-garching.mpg.de}{jstegmann@mpa-garching.mpg.de}}

\begin{abstract}
 
\end{abstract}

\begin{document}

\flushbottom
\maketitle

\begin{figure}
    \centering
    \includegraphics[width = 0.5\linewidth]{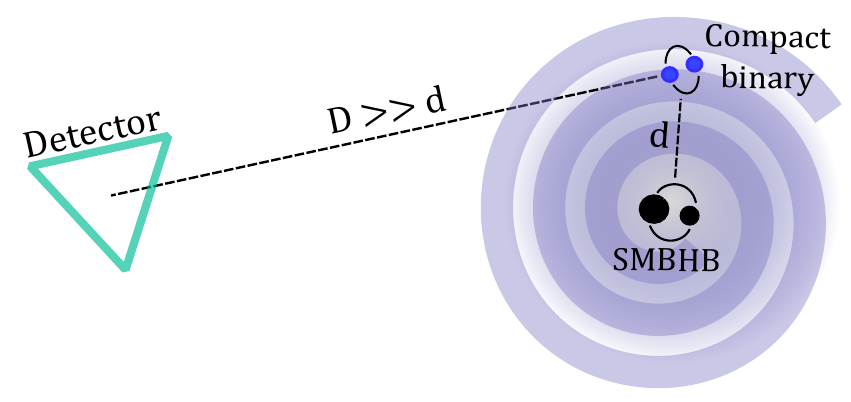}
    \caption{\textbf{Cartoon of the method proposed in this work.} The presence of an SMBHB emitting GWs causes frequency modulations in the GW emission of a compact binary source at a distance $d$. The modulations can be observed over a long observation time $T$ with proposed deci-Hz GW detectors, at a distance $D\gg d$. We show how this scenario would allow deci-Hz detectors to indirectly probe the existence of SMBHBs {within} the $\sim10^7$~--~$10^9\,\rm M_\odot$ mass range.}
    \label{fig:sketch}
\end{figure}

\vspace{-20pt}
\section*{}\label{sec:results}
\textbf{The most massive black holes in our Universe form binaries at the centre of merging galaxies. 
The recent evidence for a gravitational-wave (GW) background from pulsar timing may constitute the first observation that these supermassive black hole binaries (SMBHBs) merge. 
Yet, the most massive SMBHBs are out of reach of interferometric {GW} detectors and are exceedingly difficult to resolve individually with pulsar timing. 
These limitations call for unexplored strategies to detect individual SMBHBs in the uncharted frequency band $\lesssim10^{-5}\,\rm Hz$ in order to establish their abundance and decipher the coevolution with their host galaxies. 
Here we show that SMBHBs imprint detectable long-term modulations on GWs from stellar-mass binaries residing in the same galaxy. 
We determine that proposed deci-Hz GW interferometers sensitive to numerous stellar-mass binaries could uncover modulations from $\sim\mathscr{O}(10^{-1}$~--~$10^4)$ SMBHBs with masses {$\sim\mathscr{O}(10^7$~--~$10^8)\,\rm M_\odot$} out to redshift $z\sim3.5$. This offers a unique opportunity to map the population of SMBHBs through cosmic time, which might remain inaccessible otherwise.}

We consider the situation in which we directly detect a GW signal from a stellar-mass compact binary at luminosity distance $D$ which is accompanied by an SMBHB in its proximity at $d\ll D$, e.g., at the centre of its galaxy (see Fig.~\ref{fig:sketch}). The GWs emitted by the SMBHB perturb spacetime at the location of the compact binary which induces a small frequency modulation in the detected signal (cf. equation~\ref{eq:f-measured}):

\begin{equation}
    f_{\rm measured}(t)\approx f(t)-f(t)\zeta(t)+\mathscr{O}(\zeta^2),
\end{equation}

\noindent where $f(t)$ is the unmodulated frequency of the compact binary and $\zeta(t)\ll1$ is a small modulation. At lowest order the SMBHB causes a monochromatic modulation that can be written as a sinusoid (cf. equation~\ref{eq:zeta(t)}):

\begin{equation}
    \zeta(t)=\mathscr{A}_{\rm mod}\cos(2\pi f_{\rm mod}t+\phi_{\rm mod}),
\end{equation}

\noindent where $f_{\rm mod}$ is the frequency of the modulating GW and its amplitude $\mathscr{A}_{\rm mod}$ is of the order $\sim(\pi f_{\rm mod})^{2/3}\mathscr{M}_{\rm mod}^{5/3}/d$ with $\mathscr{M}_{\rm mod}$ being the chirp mass of the SMBHB. Thus, the detector receives modulated GWs from the compact binary (cf. equation~\ref{eq:modulated-waveform}) where the modulation is fully described by three additional physical parameters which are determined by the SMBHB: an overall amplitude $\mathscr{A}_{\rm mod}$, a modulating frequency $f_{\rm mod}$, and a phase $\phi_{\rm mod}$.

\begin{figure*}
     \centering
     \includegraphics[width=0.48\linewidth]{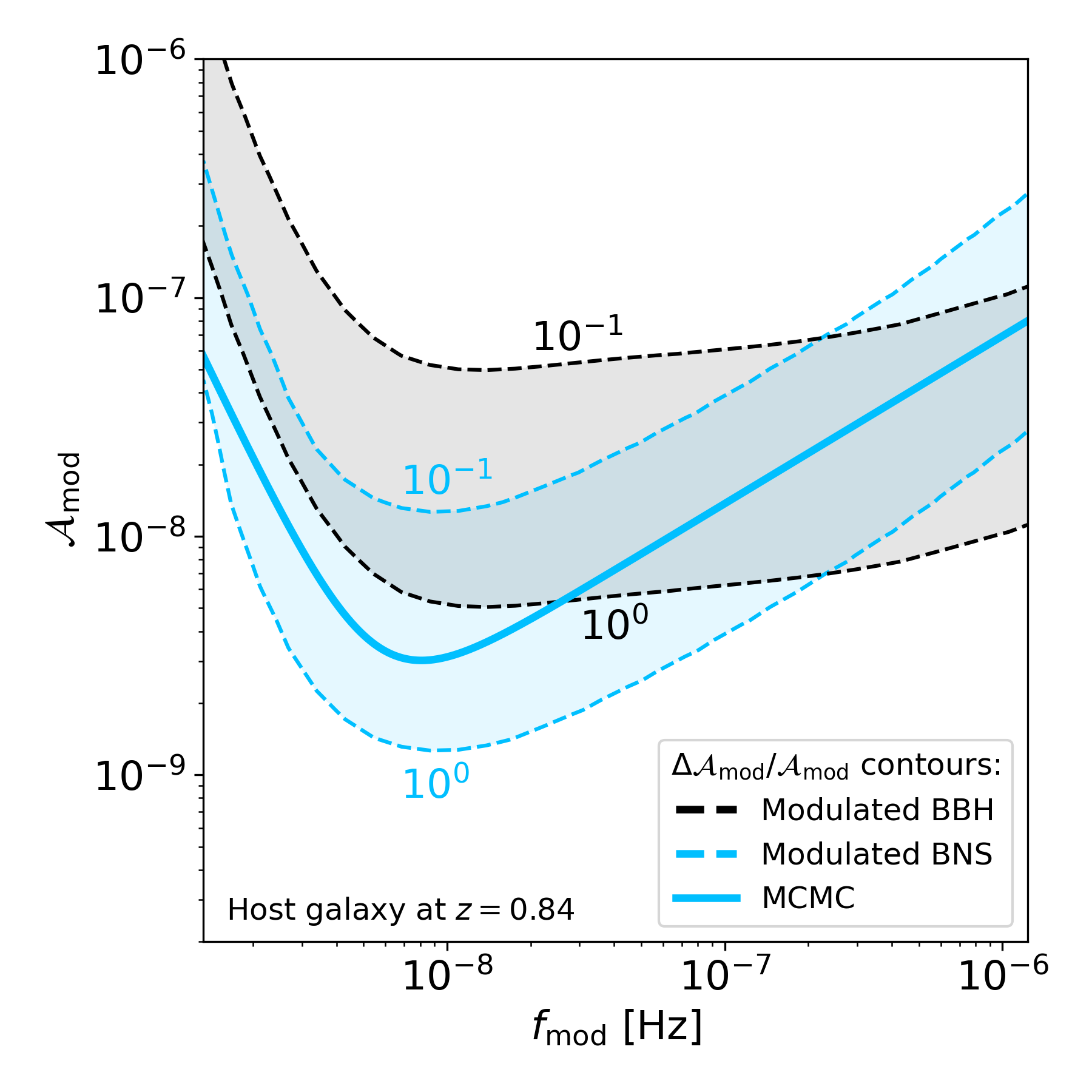}
     \includegraphics[width=0.48\linewidth]{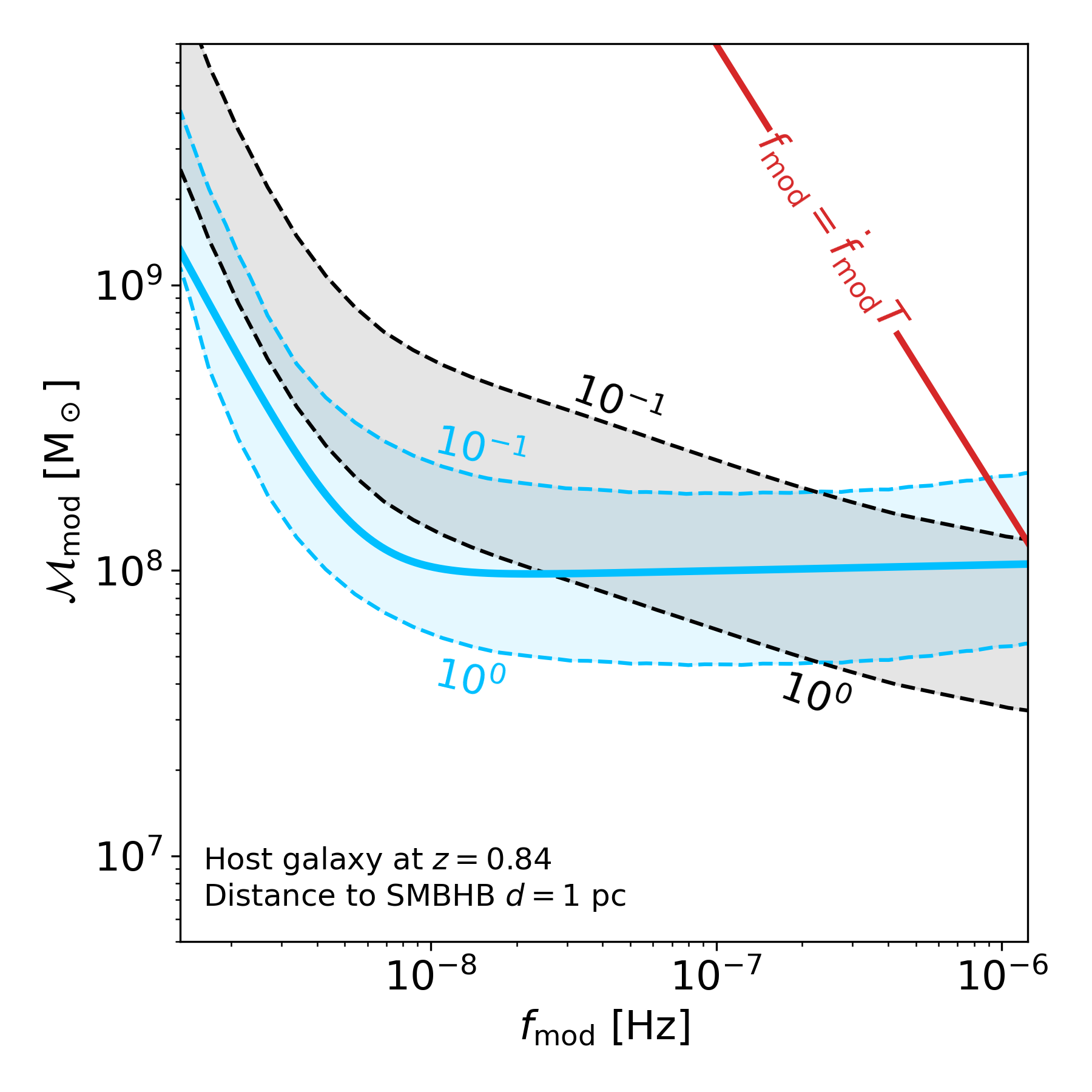}
        \caption{\textbf{Detectability contours of the modulating SMBHB amplitude $\mathscr{A}_{\rm mod}$ experienced by a stellar-mass compact binary.} The binaries are located in a fiducial host galaxy at redshift $z=0.84$. The modulated GW signal from the compact binary is observed with DECIGO over $T=10\,\rm yr$. \textit{Left panel:} To compute this sensitivity we vary the modulating amplitude $\mathscr{A}_{\rm mod}$ and frequency $f_{\rm mod}$ of the SMBHB and indicate the relative uncertainty $\Delta\mathscr{A}_{\rm mod}/\mathscr{A}_{\rm mod}$ by which a given amplitude could be recovered with {dashed} contour lines, {using parameter estimation with the Fisher matrix}. Black contour lines assume that a GW150914\cite{2016PhRvL.116x1102A}-like BBH with a chirp mass of $\mathscr{M}=28.0\,\rm M_\odot$ is observed. Blue contour lines assume a GW170817\cite{2017PhRvL.119p1101A}-like BNS ($\mathscr{M}=1.188\,\rm M_\odot$). {We also show for the BNS the sensitivity curve from a full Bayesian analysis with an MCMC (solid blue line) which corresponds to the Fisher matrix uncertainty $\Delta\mathscr{A}_{\rm mod}/\mathscr{A}_{\rm mod}\sim10^{-1/2}$ (see Methods and Fig.~\ref{fig:delays}).} \textit{Right panel:} We show the sensitivity in terms of the minimum detectable chirp mass $\mathscr{M}_{\rm mod}=(\mathscr{A}_{\rm mod}d)^{3/5}/(\pi f_{\rm mod})^{2/5}$ of the SMBHB for a fiducial distance $d=1\,\rm pc$ to the BBH/BNS. The red line indicates the frequency above which our assumption of a monochromatic SMBHB breaks down. Note that the number of SMBHBs near this limit is anyways highly suppressed (see Methods).}
        \label{fig:Amod-fmod}
\end{figure*}

{In order to investigate how well these additional parameters can be measured we employ two methods. First, we calculate the Fisher matrix of the modulated {post-Newtonian} waveform {of the compact binary which includes both higher order modes and effective spin, (see Methods)}. {The Fisher matrix approach} allows us to efficiently survey a large part of the parameter space and to estimate the expected error of fitting the parameters of the sinusoidal modulation to the noisy data of a detector (see Methods). Second, we run a series of full Bayesian parameter recovery tests employing numerical Markov chain Monte Carlo (MCMC) methods at a representative redshift of $z=0.84$. In the latter method, we explore a large grid of injected values of $\mathscr{A}_{\rm mod}$ and $f_{\rm mod}$ to find the curve delimiting {well-constrained} and {poorly-constrained} modulations, where we consider a modulation to be {well-constrained} if the one-sigma width of the posterior distribution of $\mathscr{A}_{\rm mod}$ is smaller than the injected value. {We find that this ensures that the well-constrained modulations are distinguishable from the null hypothesis of no modulation ($\mathscr{A}_{\rm mod}=0$) being present with extremely high confidence (see Methods for a more thorough discussion)}. 
Finally, as seen in Fig.~\ref{fig:Amod-fmod}, we can match the curve defined by this strict criterion to a corresponding relative error $\Delta \mathscr{A}_{\rm mod}/\mathscr{A}_{\rm mod}\equiv\sqrt{\langle(\Delta \mathscr{A}_{\rm mod})^2\rangle}/\mathscr{A}_{\rm mod}\lesssim10^{-1/2}$ that results from the simpler Fisher matrix analysis.
For the purposes of numerical efficiency, we use the scaling with redshift of the latter to extrapolate the stricter MCMC criterion.} 

Modulations caused by SMBHBs can be most easily identified with a detector that can observe a large number of compact binaries with a high signal-to-noise ratio (SNR) and over a long observation time $T$, {which is limited by the mission lifetime of the detector}. We find that these constraints single out deci-Hz GW detectors \cite{2020CQGra..37u5011A} and the large populations of binary black holes (BBHs) and binary neutron stars (BNSs) that they are expected to observe. {Thus, we adopt the sensitivity curves of the proposed space-based laser-interferometric GW detectors DECIGO and BBO \cite{2011PhRvD..83d4011Y,2017PhRvD..95j9901Y} as examples for any sufficiently sensitive detector in the deci-Hz regime.}
In Fig.~\ref{fig:Amod-fmod}, we showcase the results of our Fisher matrix {and MCMC} analysis. As an example, we assume a typical BNS and BBH located at a representative redshift $z=0.84$, and use the design sensitivity of DECIGO. The left panel shows the relative uncertainty $\Delta\mathscr{A}_{\rm mod}/\mathscr{A}_{\rm mod}$ as a function of the modulating SMBHB frequency $f_{\rm mod}$. The sensitivity curves have a characteristic shape that is similar to those of pulsar timing arrays \cite{2015CQGra..32e5004M,2019PhRvD.100j4028H}, with a peak sensitivity around $f_{\rm mod}\approx1/T$. The drop in sensitivity below $f_{\rm mod}\lesssim1/T$ reflects the fact that the observation time needs to cover at least one period of the SMBHB GW in order to establish its existence \cite{2022PhRvD.105d4005B}. Conversely, the sensitivity degrades for higher frequencies following the $f_{\rm mod}T\gg1$ limit of the hypergeometric functions in equation~\eqref{eq:hyper}. The right panel of Fig.~\ref{fig:Amod-fmod} shows the minimum detectable chirp mass $\mathscr{M}_{\rm mod}=(\mathscr{A}_{\rm mod}d)^{3/5}/(\pi f_{\rm mod})^{2/5}$ of the modulating SMBHB. Masses as low as $\mathscr{M}_{\rm mod}\sim\mathscr{O}(10^7)\,\rm M_\odot$ can be detected if the compact binaries are located at a distance $d=1\,\rm pc$ to the SMBHB, as suggested by several binary formation channels (see below). 

\begin{figure}[ht]
\centering
\includegraphics[width=0.99\linewidth]{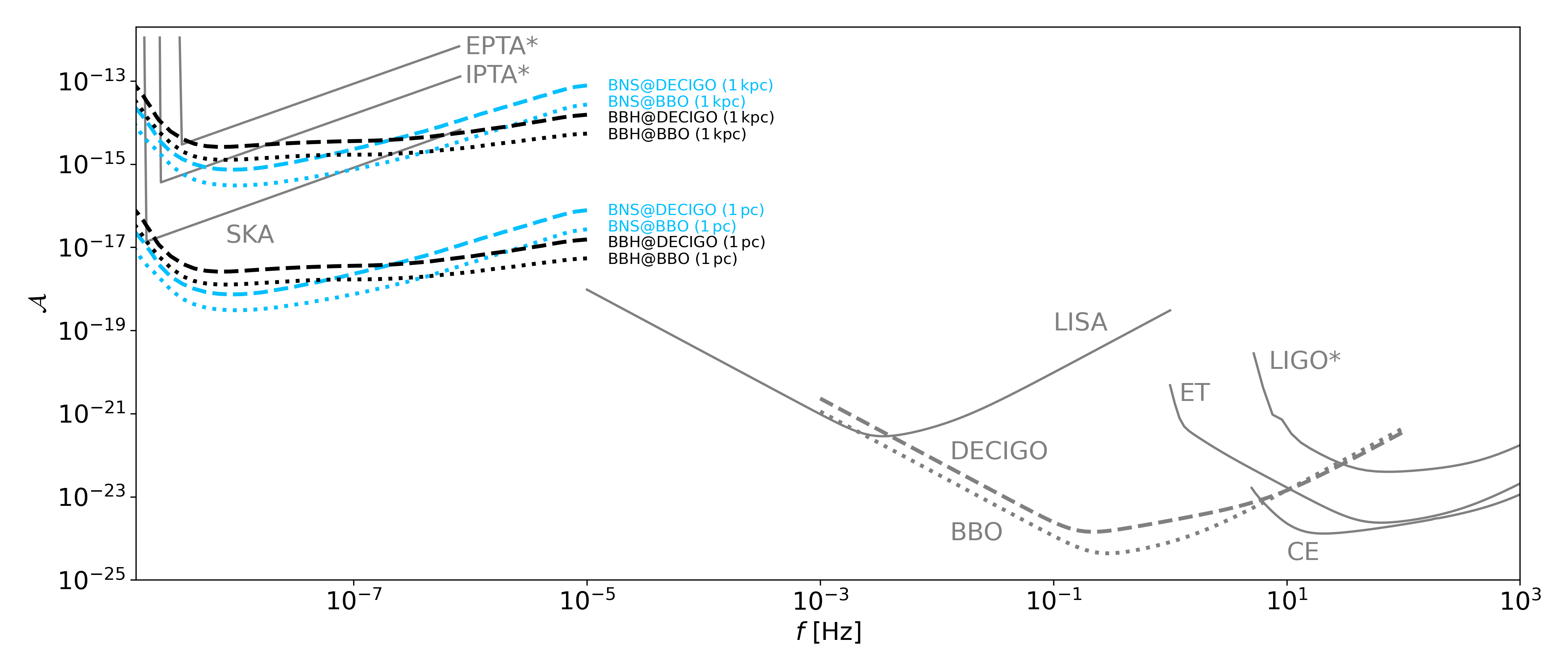}
\caption{\textbf{Landscape of different GW detectors compared to the method proposed in this work.} For each detector we plot the dimensionless strain amplitude $\mathscr{A}=\sqrt{fS_n}$ (solid lines) as function of frequency $f$ \cite{LIGOcurve,2008arXiv0810.0604H,ETcurve,CEcurve,2011PhRvD..83d4011Y,2017PhRvD..95j9901Y,2009LRR....12....2S,2015CQGra..32a5014M}, where ${}^\ast$ distinguishes currently operating detectors from planned ones. The black and blue lines show the equivalent strain sensitivity of our method, which corresponds to the GW strain amplitude the SMBHB would produce on Earth if it were \textit{indirectly detected} (with $\Delta \mathscr{A}_{\rm{mod}}/\mathscr{A}_{\rm{mod}}={10^{-1/2}}$) through the modulations of a \textit{directly detected} signal from a typical BBH (e.g., GW150914\cite{2016PhRvL.116x1102A}) and a typical BNS (e.g., GW170817\cite{2017PhRvL.119p1101A}), respectively. Dashed lines correspond to an observation with DECIGO and dotted lines to an observation with BBO. We emphasise that this method can only detect SMBHBs which are concurrent with a compact object merger in its proximity at a given separation $d$ which limits the number of potential sources (see main text). Nevertheless, the resulting sensitivity can greatly outperform other nano-Hz observatories, such as EPTA, IPTA, and SKA.}
\label{fig:A-fmod}
\end{figure}

In Fig.~\ref{fig:A-fmod}, we compare the sensitivity of our method to other currently operating and planned GW detectors. Since the SMBHBs perturb compact binaries in their proximity, we consider an equivalent strain sensitivity that corresponds to the strain amplitude the distant SMBHBs would produce on Earth. The resulting equivalent sensitivities of DECIGO/BBO in the nano-Hz band are comparable to or better than those of current and planned pulsar timing arrays. For distances between the stellar-mass compact binary and the SMBHB of $d=1\,\rm pc$, suggested by various compact object binary formation channels (see below), the obtained sensitivity would outperform the anticipated SKA sensitivity by two orders of magnitude. However, we stress that this striking sensitivity only applies to SMBHBs which are accompanied by a compact binary in their proximity, inherently limiting the number of detectable systems. The sensitivity differences between BNSs and BBHs are caused by their different masses and frequency evolution (see Methods). They are negligible for the purpose of estimating the detection rate (see below). Thus, we will henceforth refer to both as generic compact binary sources.

\begin{figure}[t!]
    \centering
    \includegraphics[width =0.9\columnwidth]{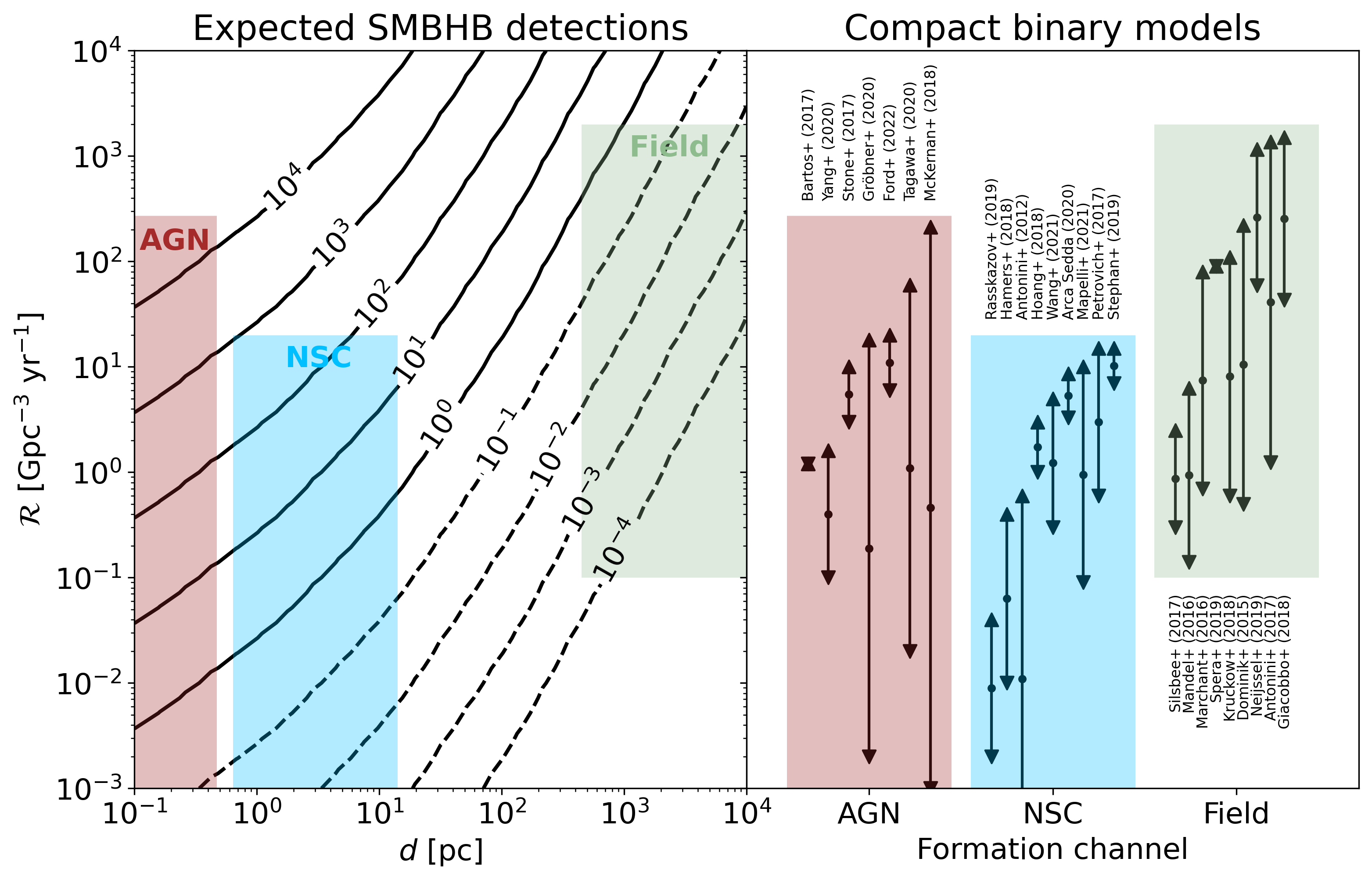}
    \includegraphics[width = 0.89\columnwidth]{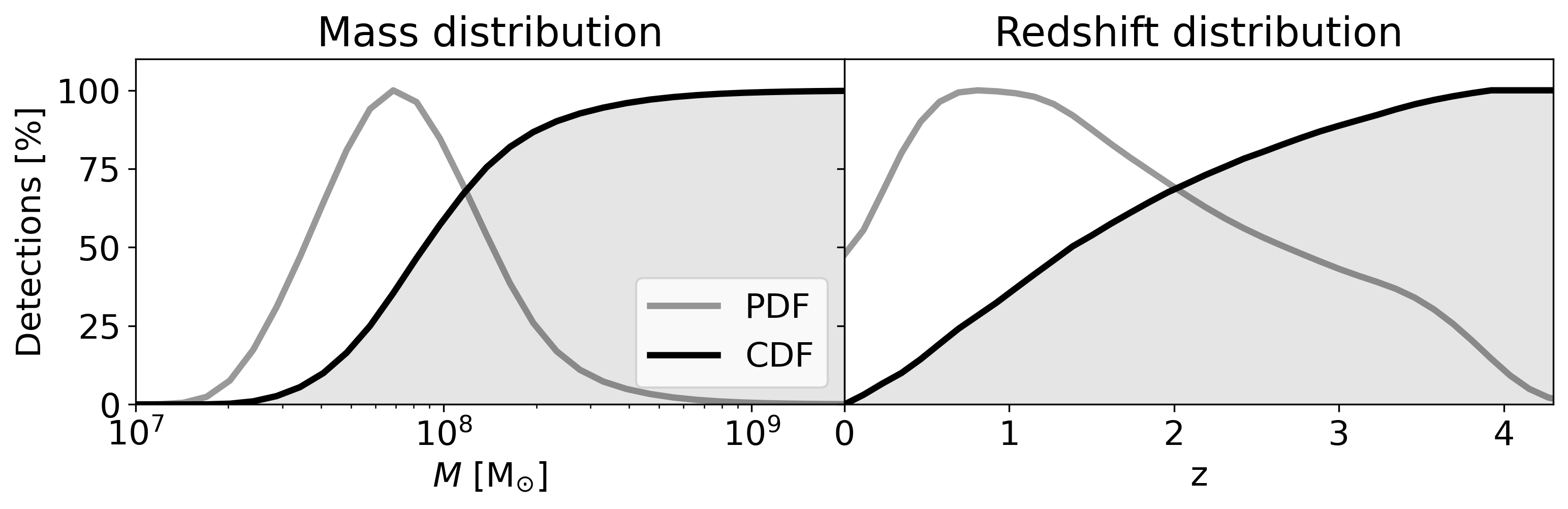}
    \caption{\textbf{Total number and distribution of the expected SMBHB detections over a $T=10\,\rm yr$ observation with a deci-Hz detector.} \textit{Upper panels:} We show the expected number of SMBHB detections as a function of the compact binary merger rate $\mathscr{R}$ and the distance $d$ from the Galactic centre at which the binaries typically merge. Compact binary mergers can occur through one or more formation channels. Rate predictions by several AGN \cite{Bartos_2017,2017MNRAS.464..946S,2018ApJ...866...66M,2020A&A...638A.119G,2020ApJ...896..138Y,2020ApJ...898...25T,2022MNRAS.517.5827F}, NSC \cite{2012ApJ...757...27A,2017ApJ...846..146P,2018ApJ...856..140H,2018ApJ...865....2H,2019MNRAS.488...47F,2019ApJ...878...58S,2019ApJ...881...20R,2020ApJ...891...47A,2021MNRAS.505..339M,2021ApJ...917...76W}, and field channel models \cite{2015ApJ...806..263D,2016MNRAS.458.2634M,2016A&A...588A..50M,2017ApJ...841...77A,2017ApJ...836...39S,2018MNRAS.481.1908K,2018MNRAS.480.2011G,2019MNRAS.490.3740N,2019MNRAS.485..889S} are shown by coloured shaded areas and further detailed by the errorbars in the upper right panel. We assume a plausible range of $d \lesssim \mathscr{O}(10^{-1})\,\rm pc$, $\sim \mathscr{O}(10^0)\,\rm pc$, and $\sim \mathscr{O}(10^3)\,\rm pc$ to represent the typical half mass radii of AGN discs, NSCs and large galaxies, respectively. Crucially, the large majority of models for the AGN and NSC formation channels would guarantee tens to hundreds of SMBHB detections. \textit{Lower panels:} We show the mass and redshift distributions of the detectable SMBHBs. The method is sensitive to individual sources with total mass {$M\sim10^7$~--~$10^8\,\rm M_\odot$} up to redshifts of $z\lesssim 3.5$.}
    \label{fig:RBNS-d}
\end{figure}

Given the sensitivity curves above, we estimate the expected number and properties of SMBHBs that cause detectable modulations on the GWs from compact binaries observed with deci-Hz instruments. For this purpose, we adopt an SMBHB population model based on the Millennium Simulation\cite{2010ApJ...718.1400F}. 
We then distribute compact binaries within galaxies following the latter's total stellar mass and evaluate the probability that they coincide with an SMBHB that produces a detectable modulation (see Methods). 

Fig.~\ref{fig:RBNS-d} shows that the distribution of detectable SMBHBs strongly peaks {within} the range of $10^7$ to $10^9\,\rm M_\odot$, which represents the best compromise between the total quantity of available SMBHBs and the strength of the modulation they produce. {Most detections are limited to relatively low redshifts of $0.5\lesssim z\lesssim1$. However, a significant fraction of the potential detections is distributed in the large cosmological volume enclosed between $z\sim 1$ and $z \sim 3.5$. We highlight that for the majority of detected compact binaries the expected distance measurement and angular resolution of DECIGO/BBO would be sufficient to identify the host galaxy of the SMBHB \cite{PhysRevD.80.104009}, allowing for targeted multi-messenger searches of sub-pc SMBHBs. This shows how our method can unlock individual GW-based detections of the most massive SMBHBs in our Universe complementing current pulsar timing arrays and the upcoming LISA detector \cite{2017arXiv170200786A} at higher redshifts and wider SMBHB separations, respectively.}

Fig.~\ref{fig:RBNS-d} also shows the expected total number of detectable SMBHBs as a function of the distance $d$ and the compact binary merger rate $\mathscr{R}$. The distance $d$ at which the compact binaries typically reside from the centre of their galaxy is a consequence of how it was formed; multiple formation channels have been proposed and are being {actively} discussed in literature in order to explain the origin of the compact binary mergers observed with LIGO-Virgo \cite{2022LRR....25....1M}. For instance, in the field scenario compact binaries are formed far away from their galactic centre from the evolution of isolated massive stellar binaries, triples, or more complex stellar systems \cite{2015ApJ...806..263D,2016MNRAS.458.2634M,2016A&A...588A..50M,2017ApJ...841...77A,2017ApJ...836...39S,2018MNRAS.481.1908K,2018MNRAS.480.2011G,2019MNRAS.490.3740N,2019MNRAS.485..889S}. In contrast, binary mergers near the galactic centre could be promoted by strong environmental effects, in e.g., nuclear star clusters (NSCs) \cite{2012ApJ...757...27A,2017ApJ...846..146P,2018ApJ...856..140H,2018ApJ...865....2H,2019MNRAS.488...47F,2019ApJ...878...58S,2019ApJ...881...20R,2020ApJ...891...47A,2021MNRAS.505..339M,2021ApJ...917...76W} or inside the disks of active galactic nuclei (AGNs) \cite{Bartos_2017,2017MNRAS.464..946S,2018ApJ...866...66M,2020A&A...638A.119G,2020ApJ...896..138Y,2020ApJ...898...25T,2022MNRAS.517.5827F}. While the latter have only been investigated for galactic centres that harbour a single supermassive black hole (SMBH), it is plausible to assume that they work similarly in galaxies which host SMBHBs. Given the current LIGO-Virgo observations, it is uncertain which of these or other formation channels (e.g., in globular clusters or young open star clusters) dominate the merger rate in the Universe, or whether multiple formation pathways coexist \cite{2021ApJ...910..152Z}. 

The majority of current NSC models yield a merger contribution of $\mathscr{R}\sim\mathscr{O}(10^{-1}$~--~$10^{1})\,\rm Gpc^{-3}\,yr^{-1}$, and NSCs are observed to have a typical half-mass radius of $\sim\mathscr{O}(10^0)\,\rm pc$. As shown in Fig.~\ref{fig:RBNS-d}, this would guarantee $N_{\rm det}\sim\mathscr{O}(10)$ detections of SMBHBs after $T=10\,\rm yr$ of observation. In the AGN channel, compact binaries merge even closer to the galactic centre, as they are within the extent $d\lesssim\mathscr{O}(10^{-1})\,\rm pc$ of the AGN disk. Thus, we find that current AGN models allow for $N_{\rm det}\sim\mathscr{O}(10^{-1}$~--~$10^{4})$ SMBHB detections. Note that our method allows SMBHB detections even if these channels only amount to a subdominant fraction of the total rate of compact binary mergers \cite{2023PhRvX..13a1048A}. Only in the least favourable scenario, in which all compact binaries are exclusively formed in the galactic field, do we expect no detections. In that case, we can assume the compact binaries to follow the galactic stellar mass distribution. Then, $d$ is similar to the typical half-mass radius $\sim\mathscr{O}(10^3)\,\rm pc$ of massive galaxies, yielding $N_{\rm det}\lesssim\mathscr{O}(10^{-1})$ for $\mathscr{R}\lesssim\mathscr{O}(10^{1}$~--~$10^{3})\,\rm Gpc^{-3}\,yr^{-1}$. {Hence, it is plausible that a significant number of compact binaries experience a detectable modulation due to the GWs from central SMBHBs, unless all of them merge in the galactic field.}

Finally, our method can be used to distinguish different compact binary formation channels. For instance, the observation of a modulated compact binary would suggest a formation within a galactic nucleus, whereas the absence of such detection places strong upper limits of $\mathscr{R}\lesssim\mathscr{O}(10^{-3})\,\rm Gpc^{-3}\,yr^{-1}$ (AGN) and  $\lesssim\mathscr{O}(10^{-2})\,\rm Gpc^{-3}\,yr^{-1}$ (NSC) on the rate of compact binary mergers that originate from the galactic centres.

{It is important to note that we have restricted our analysis to sources that exhibit at least one entire sinusoidal modulation. In this way, we exclude the possibility that the perturbation to the GW may be degenerate with any additional slowly accumulating phase shift. For instance, GWs from a compact binary orbiting a central (single or binary) SMBH may exhibit a detectable phase shift caused by Doppler motion if $d\lesssim\mathscr{O}(1)\,\rm pc$ \cite{2018CQGra..35e4004M}. This Doppler modulation may only be periodic, and thus degenerate with our effect, if the observational time is comparable or longer than the orbital period of the binary around the SMBH(B)s. For $T=10\,\rm yr$ and $M=10^8\,\rm M_\odot$, that is the case only if $d\lesssim\mathscr{O}(10^{-2})\,\rm pc$. Such small separations arise exclusively in formation channels within migration traps of AGN disks \cite{2016ApJ...819L..17B,2022PhRvD.106f4056S}. {Conversely, the simultaneous detection of our effect and the acceleration of the compact binary\cite{2020PhRvD.101f3002T} in the vicinity of a central SMBHBs would be an unambiguous signature of its existence.} Furthermore, the periodicity of the sinusoidal modulation occurs on nHz frequencies. While strong gravity effects, such as apsidal and spin precession may produce some oscillatory phase modulations, they {occur on frequencies which are several orders of magnitude higher for binaries in the deci-Hz band. Additionally, the latter} is evolving throughout the chirp, and can therefore not be degenerate with a constant low frequency modulation (see also Fig.~\ref{fig:MCMC}, where we tested the degeneracy explicitly with the effective spin parameter).}

To conclude, we have shown that the inherent sensitivity of proposed laser interferometers may be exploited to detect individual SMBHBs in the previously inaccessible mass range $\gtrsim 10^7$ M$_{\odot}$, at redshifts beyond the capacity of both current and future pulsar timing arrays. Currently, there are no other proposed methods that would realistically guarantee a direct detection of GWs from individual SMBHBs in the nano-Hz band. A deci-Hz GW detector has exceptional scientific potential and should therefore be pursued with urgency, as it would open a large volume of the Universe to GW-based observations, simultaneously in both the deci-Hz and the nano-Hz band.

\section*{Methods\footnote{Throughout this work we set $G=c=1$. If not stated differently, the magnitude and unit vector of some spatial vector $\mathbf{V}$ are written as $V=\left|\mathbf{V}\right|$ and $\mathbf{\hat{V}}=\mathbf{V}/V$, respectively.}}\label{sec:methods}

\subsection*{Parameter estimation}\label{sec:parameter-estimation} 
A standard quantity to estimate the detectability of a GW strain $h$ with a given detector is the SNR $\rho$ defined as 

\begin{equation}
    \rho[h]=\sqrt{\left(h|h\right)},
\end{equation}

\noindent where we defined the noise-weighted inner product

\begin{equation}\label{eq:noise-weighted}
    \left(a|b\right)=2\int_{f_{\rm min}}^{f_{\rm max}}{\rm d} f\,\frac{\Tilde{a}(f)\Tilde{b}^\ast(f)+\Tilde{b}(f)\Tilde{a}^\ast(f)}{S_n(f)}.
\end{equation}

\noindent Here, $S_n(f)$ is the noise power spectral density of the detector and the integration boundaries $f_{\rm min}$ and $f_{\rm max}$ correspond to the signal frequencies at the beginning and the end of the observation, respectively. In general, the waveform $h$ depends on a set of physical parameters $h=h(\theta^0,\theta^1,\theta^2,\dots)$. 
Let $\hat{\theta}^i$ be the correct values of the signal and $\hat{\theta}^i+\Delta\theta^i$ the best-fit parameters for a given realisation of the noise. For a large SNR $\rho$ the measurement uncertainty $\Delta\theta^i$ follows a Gaussian distribution \cite{1992PhRvD..46.5236F,1994PhRvD..49.2658C,2008PhRvD..77d2001V}

\begin{equation}
    p(\Delta\theta^i)\propto\exp\left(-\frac{1}{2}\Gamma_{ij}\Delta\theta^i\Delta\theta^j\right),
\end{equation}

\noindent where $\Gamma_{ij}$ is the Fisher information matrix defined as

\begin{equation}
    \Gamma_{ij}=\left(\frac{\partial h}{\partial\theta^i}\Bigg|\frac{\partial h}{\partial\theta^j}\right).
\end{equation}

\noindent Thus, the root-mean-square error of $\theta^i$ is

\begin{equation}
    \sqrt{\langle(\Delta \theta^i)^2\rangle}=\sqrt{\Sigma^{ii}},
\end{equation}

\noindent where $\Sigma=\Gamma^{-1}$ is the inverse matrix of $\Gamma$. Thus, the relative uncertainty of the $i$-th parameter can be estimated as $\sqrt{\Sigma^{ii}}/\hat{\theta}^i$.

\begin{figure}[t!]
    \centering
    \includegraphics[width = 0.99\columnwidth]{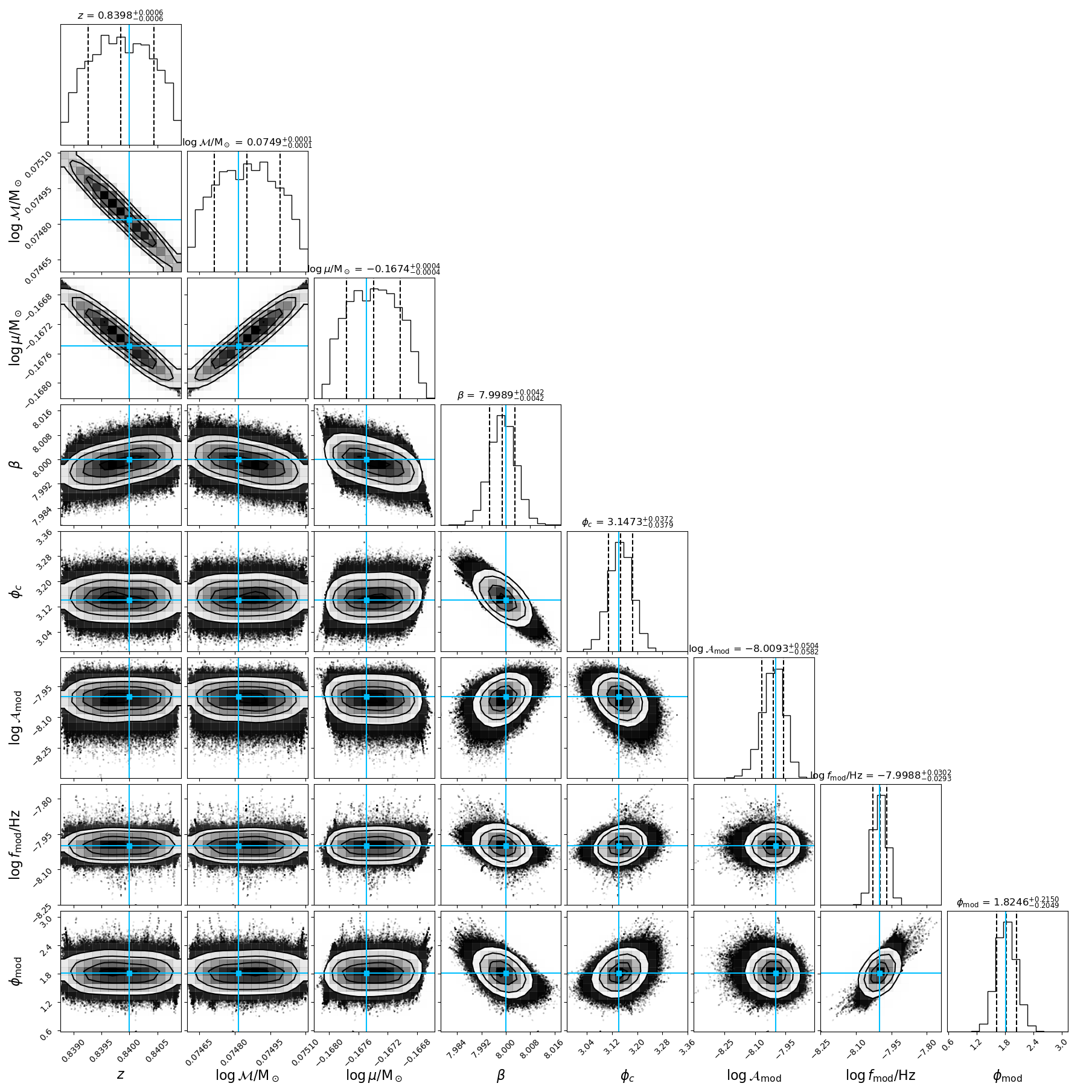}
    \caption{\textbf{Parameter estimation with a Markov chain Monte Carlo (MCMC) method of a modulated carrier GW.} We assumed a BNS at redshift $z=0.84$ (consistent with Fig.~\ref{fig:Amod-fmod}), with chirp mass $\mathscr{M}=1.188\,\rm M_\odot$, reduced mass $\mu=0.68\,\rm M_\odot$, effective spin parameter $\beta=8.0$ \cite{1994PhRvD..49.2658C}, and phase $\phi_c=3.14$ and a modulation with amplitude $\mathscr{A}_{\rm mod}=10^{-8}$, frequency $f_{\rm mod}=10^{-8}\,\rm Hz$, and phase $\phi_{\rm mod}=1.81$ (shown in blue). {We assume that the BNS is observed with DECIGO ($T=10\,\rm yr$) which results in an SNR of the carrier signal of $\rho\approx175$. The posterior samples of the waveform parameters are obtained with} the affine-invariant ensemble sampler {\tt emcee} \cite{emcee} for MCMC proposed by Ref. \cite{Goodman2010} for a chain length of $10^6$, $40$ walkers, and a discarded burn-in length of $5\times10^4$. The dashed lines indicate the 0.16, 0.50, and 0.84 percentile of the posteriors, which for this system agree well with the injected ground truth values. This plot was generated using the {\tt python} package {\tt corner}\cite{corner}.}
    \label{fig:MCMC}
\end{figure}

{
The limitations of the Fisher matrix formalisms have been thoroughly explored in Ref. \cite{2008PhRvD..77d2001V}. It is generally understood that, when it comes to GW parameter estimation, reliable estimates often require the full sampling of parameter posteriors in order to test for degeneracies and other features of the low SNR regime. In GW parameter estimation, the goal is to maximise a likelihood $\mathcal{L}$ of the form}
\begin{align}
    \mathcal{L}\left(\Theta \right) \propto \exp\left[ - \big(h(\Theta) - h(\Theta_{\rm inj}) \lvert h(\Theta) - h(\Theta_{\rm inj}) \big) \right],
\end{align}
{where $\Theta_{\rm inj}$ represent the parameters of an injected signal. In this work we complement our Fisher matrix estimates with a series of MCMC based parameter recovery tests. We then adjust the Fisher matrix detectability threshold in order to reflect the more conservative MCMC-based results. To explore the posterior distribution functions for the parameters $\Theta$, we use the affine-invariant ensemble sampler {\tt emcee} \cite{emcee} for MCMC proposed by Ref. \cite{Goodman2010} and perform several tests with post-Newtonian waveform models. {In the MCMCs we use uniform priors for all waveform parameters, except $\mathscr{A}_{\rm mod}$ and $f_{\rm mod}$ for which we impose log-uniform priors between the wide ranges $10^{-12}$~--~$10^{-4}$ and $10^{-12}$~--~$10^{-4}\,\rm Hz$, respectively.} A realisation of an MCMC test for a detectable modulation can be seen in Fig.~\ref{fig:MCMC}. In Fig. \ref{fig:delays} we show the sensitivity curve that results from injecting a grid of values of $\mathscr{A}_{\rm mod}$ and $f_{\rm mod}$. This sensitivity curve is used in Fig.~\ref{fig:Amod-fmod}.}

\subsection*{Unmodulated Waveform}\label{sec:Newtonian-waveform}
The simplest waveform from a circular binary can be obtained by modelling it as two Newtonian point masses $m_1$ and $m_2$ whose orbital frequency grows secularly due to the energy-loss by GWs \cite{1994PhRvD..49.2658C}:

\begin{equation}
    \Tilde{h}_{\rm Newton}(f)=\frac{Q}{D}\mathscr{M}^{5/6}f^{-7/6}\exp[i\Psi(f)],\label{eq:unmodulated-waveform}
\end{equation}

\noindent where $f\geq0$, $D$ is the luminosity distance to the binary, $Q$ is a numerical factor accounting for the angular emission pattern of the source and the antenna response of the detector to the GW, and there is a phase

\begin{equation}
    \Psi(f)=2\pi ft_c-\phi_c-\frac{\pi}{4}+\frac{3}{4}\left(8\pi\mathscr{M}f\right)^{-5/3}.
\end{equation}

\noindent Thus, this waveform can be written in terms of four physical parameters: an overall amplitude $\mathscr{A}=Q\mathscr{M}^{5/6}/D$, a chirp mass $\mathscr{M}=(m_1m_2)^{3/5}/(m_1+m_2)^{1/5}$, a collision time $t_c$ describing the point in time of the merger, and a constant phase $\phi_c$. If the binary is at a cosmological distance at redshift $z$, we need to replace its chirp mass by $\mathscr{M}\rightarrow\mathscr{M}(1+z)$.

{The advantage of Newtonian waveforms is that they may be easily treated in the Fisher matrix formalism, and that the scalings and degeneracies are well understood \cite{1994PhRvD..49.2658C}. However, Newtonian waveforms do not suffice for the requirements of parameter estimation already in current GW data analysis pipelines. In order to surpass these limitations, we include post-Newtonian (PN) corrections for our Fisher matrix estimate and the MCMC based numerical tests. We include the full phase evolution for non-spinning binaries up to 3 PN, and include the effect of spin through an additional phase modification at 1.5 PN order. The coefficients of the GW phase are taken from Ref. \cite{2014blanchet}. We also include the effect of all higher order modes up to powers of $x^{2.5}$, where $x$ is the PN parameter, also taken from Ref. \cite{2014blanchet}. Effectively, this introduces also a dependency of the waveform on the reduced mass $\mu=m_1m_2/(m_1+m_2)$ and the effective spin parameter $\beta$.}

\subsection*{Modulated Waveform}\label{sec:Modulated-waveform}
We consider the situation where we have a so-called "carrier" source that emits GWs within the frequency bandwidth of our detector, and another "background" source radiating at much lower frequencies. The background GW modulates the apparent frequency of the carrier GW by perturbing spacetime along the detector-carrier line-of-sight. Thus, the \textit{measured} frequency can be written as \cite{MaggioreMichele2018GWV2}

\begin{equation}
    f_{\rm measured}(t)=\frac{1}{T(t)+\Delta T(t)}\approx f(t)-f(t)\zeta(t)+\mathscr{O}(\zeta^2),\label{eq:f-measured}
\end{equation}

\noindent where we substituted the unmodulated frequency $f(t)\equiv1/T(t)$ and a small modulation $\zeta(t)=\Delta T(t)/T(t)\ll1$. Note that the modulating effect of the background GW applies to any generic periodic carrier signal. For instance, using telescopes to search for modulations of the rotational period of millisecond pulsars is the aim of large-scale observational campaigns to detect low-frequency GWs \cite{1979ApJ...234.1100D,1978SvA....22...36S,1982MNRAS.199..659M,1983MNRAS.203..945B,1983ApJ...265L..39H,1990ApJ...361..300F,1994ApJ...428..713K,2005ApJ...625L.123J,2006ApJ...653.1571J,2009MNRAS.394.1945H}. In this work, we assume that the carrier source is a binary emitting GWs (see above) whose frequency evolves as \cite{1963PhRv..131..435P}

\begin{equation}
    f(t)=\frac{1}{8\pi}\left[\frac{1}{5}(t_c-t)\mathscr{M}^{5/3}\right]^{-3/8}\label{eq:f(t)}.
\end{equation}

\noindent The modulation is given by \cite{MaggioreMichele2018GWV2}

\begin{equation}
    \zeta(t)=\frac{D^iD^j}{2(1+\hat{\mathbf{b}}\cdot\hat{\mathbf{D}})}h^{\rm TT}_{ij}(t,\mathbf{x}=0)
    -\frac{D^iD^j}{2(1+\hat{\mathbf{d}}\cdot\hat{\mathbf{D}})}h^{\rm TT}_{ij}(t-D,\mathbf{x}=\mathbf{D}),\label{eq:z}
\end{equation}

\noindent where $\mathbf{x}=0$ is the position of the Earth, $\mathbf{x}=\mathbf{D}$ the position of the carrier, $\mathbf{x}=\mathbf{b}$ the position of the background source, $\mathbf{d}=\mathbf{b}-\mathbf{D}$ the vector from the background source to the carrier, and $h^{\rm TT}$ the linear spacetime metric perturbation in the transverse-traceless gauge which is induced by the background source. {Eq.~\eqref{eq:z} is identical to the formalism used to describe the GW-induced modulation of a pulsar\cite{MaggioreMichele2018GWV2}, except that in that case Earth and pulsar are considered much closer than the SMBHB ($D\ll d$) so that the GW is incident from the same direction, i.e., $\hat{\mathbf{b}}\approx \hat{\mathbf{d}}$.}

In this work, we are considering the {opposite} case where the carrier and the background source are close to each other, but far away from Earth, i.e., $D\gg d$ {and $\hat{\mathbf{b}}\neq\hat{\mathbf{d}}$}. In pulsar timing literature, the first term on the r.h.s. of equation~\eqref{eq:z} is usually referred to as Earth term and the second to as pulsar term. {The amplitudes of the first and second term on the r.h.s. of equation~\eqref{eq:z} decrease with distance $D$ and $d$, respectively.} Due to the large distance to the background source we can therefore neglect the former in our work, and focus on the spacetime distortion the background source produces at the location of the carrier. If the background source is an SMBHB with a chirp mass $\mathscr{M}_{\rm mod}$ that emits monochromatic GWs at a frequency $f_{\rm mod}$ the distant modulation term can be written as \cite{MaggioreMichele2008GWV1,MaggioreMichele2018GWV2}

\begin{equation}
    \zeta(t)=\mathscr{A}_{\rm mod}\cos(2\pi f_{\rm mod}t+\phi_{\rm mod}),\label{eq:zeta(t)}
\end{equation}

\noindent where its amplitude is of the order $\mathscr{A}_{\rm mod}\approx(\pi f_{\rm mod})^{2/3}\mathscr{M}_{\rm mod}^{5/3}/d$. Here, we averaged over a random orientation of $\mathbf{d}$ w.r.t. $\mathbf{D}$ and a random  inclination of the orbital plane of the SMBHB. In time-domain the GW strain from the carrier is given by \cite{1994PhRvD..49.2658C}

\begin{equation}\label{eq:h-t}
    h(t)=\frac{(384/5)^{1/2}\pi^{2/3}Q\mu M}{Dr(t)}\cos{\left(2\pi\int^t  f_{\rm measured}(t')\,{\rm d}t'\right)},
\end{equation}

\noindent where $M=m_1+m_2$ and $\mu=m_1m_2/M$ are the total and reduced mass, respectively, and the orbital separation of the binary is

\begin{equation}
    r(t)=\left(\frac{256}{5}\mu M^2\right)^{1/4}\left(t_c-t\right)^{1/4}.
\end{equation}

\noindent Considering the integral in equation~\eqref{eq:h-t} we have

\begin{align}
    \phi(t)&\equiv2\pi\int^t  f_{\rm measured}(t')\,{\rm d}t'\approx 2\pi\int^t f(t')\,{\rm d}t'-2\pi\int^t f(t')\zeta(t')\,{\rm d}t',
\end{align}

\noindent where the first integral on the r.h.s. evaluates to \cite{1994PhRvD..49.2658C}

\begin{equation}
    2\pi\int^t f(t')\,{\rm d}t'=-2\left(\frac{T}{5\mathscr{M}}\right)^{5/8}+\phi_c,
\end{equation}

\noindent where $T=t_c-t$ is the time-to-coalescence
and the second integral becomes

\begin{align}
    \label{eq:hyper}
    &2\pi\int^t f(t')\zeta(t')\,{\rm d}t'=-\frac{2\mathscr{A}_{\rm mod} T}{13\times5^{5/8}\left(T \mathscr{M}^{5/3}\right)^{3/8}}\nonumber\\
    &\times\Biggl[10\pi f_{\rm mod}T\sin(\phi_{\rm mod})\,{}_1F_2\left(\frac{13}{16};\frac{3}{2},\frac{29}{16};-f_{\rm mod}^2T^2\pi^2\right)+13\cos(\phi_{\rm mod})\,{}_1F_2\left(\frac{5}{16};\frac{1}{2},\frac{21}{16};-f_{\rm mod}^2 T^2\pi^2\right)\Biggr]+\phi_{c,\rm mod}.
\end{align}

\noindent The hypergeometric functions ${}_1F_2$ evaluate to one in the limit $(f_{\rm mod}T)\rightarrow0$ where the period of the modulating GW is much longer than the observation time, and they are zero in the opposite regime $(f_{\rm mod}T)\rightarrow\infty$. Next we consider the Fourier transform:

\begin{equation}
    \Tilde{h}(f)=\int_{-\infty}^\infty h(t)e^{2\pi ift}\,{\rm d}t.
\end{equation}

\noindent In the stationary-phase approximation we get \cite{1994PhRvD..49.2658C}:

\begin{equation}
    \Tilde{h}(f)=\frac{Q}{D}\mathscr{M}^{5/6}f^{-7/6}\exp[i(2\pi ft-\phi(f)-\pi/4)],\label{eq:modulated-waveform}
\end{equation}
where $\phi(f)\equiv\phi(f(T))$ and

\begin{align}
    T(f)&=5(8\pi f)^{-8/3}\mathscr{M}^{-5/3}\label{eq:T}.
\end{align}

\noindent Note that equation~\eqref{eq:modulated-waveform} reduces to equation~\eqref{eq:unmodulated-waveform} if there is no modulation $\mathscr{A}_{\rm mod}=0$. In general, the modulated waveform in equation~\eqref{eq:modulated-waveform} now depends on four \textit{additional} physical parameters: an overall amplitude $\mathscr{A}_{\rm mod}$, a modulating frequency $f_{\rm mod}$, and the phases $\phi_{\rm mod}$ and $\phi_{\rm c,mod}$. However, we can further reduce the degrees of freedom to seven in total by omitting $\phi_{\rm c,mod}$ through an appropriate redefinition of $\phi_c\rightarrow\phi_c+\phi_{c,\rm mod}$. {When we calculate the noise-weighted inner product in Eq.~\eqref{eq:noise-weighted} the integration limits are chosen to be $f_{\rm min}=f(T=10\,{\rm yr})$ and $f_{\rm max}=f_{\rm ISCO}\equiv 1/12\pi\sqrt{6}(m_1+m_2)$, i.e., we assume that we observe all binaries during their last $10\,\rm yr$ before they merge. In reality, when a detector begins to observe the binaries will be distributed uniformly in $T$, i.e., some of the binaries are in fact observed for less than $10\,\rm yr$ before they merge ($f_{\rm min}>f(T=10\,{\rm yr})$). Binaries which are observed for less than $10\,\rm yr$ are less likely to exhibit one full cycle of a low-frequency modulation, meaning that the modulation is more liable to be degenerate with other environmental effects (see main text). However, for the typical carrier sources we are considering, more than $\sim90\,\%$ of the SNR is accumulated in the last year before merger. Indeed, we find that the bound on $\Delta \mathscr{A}_{\rm mod}/\mathscr{A}_{\rm mod}$ converges to the results shown in Fig.~\ref{fig:Amod-fmod} well within a factor of two for sources that are observed for $\sim3\,\rm yr$ or more. Thus, our population estimates will at most be reduced by $\sim30\,\%$ by this consideration.}

\subsection*{Carrier populations}\label{sec:carrier-populations}
During the first three LIGO-Virgo observing runs the GWs of ninety mergers of binary black holes (BBHs), binary neutron stars (BNSs), and neutron star-black hole (NSBH) binaries haven been directly detected \cite{2019PhRvX...9c1040A,2019ApJ...882L..24A,2021PhRvX..11b1053A,2021arXiv211103606T,2023PhRvX..13a1048A}. From this sample it is possible to reconstruct the merger rate and parameter distribution of the underlying astrophysical population of compact binaries \cite{2023PhRvX..13a1048A}.
Owing to their stronger signal, most mergers have been detected of BBHs. Their inferred merger rate in the local Universe is \cite{2023PhRvX..13a1048A}

\begin{equation}
    \mathscr{R}_{\rm BBH}=17.9\,\mathrm{-}\,44\,\rm Gpc^{-3}\,yr^{-1}\quad(90\%\,\rm C.I.),
\end{equation}

\noindent at a fiducial redshift $z=0.2$. The rate is observed to scale with redshift as $(1+z)^\kappa$ where $\kappa=2.9^{+1.7}_{-1.8}$ for $z\lesssim1$. It is plausible to assume that this trend extends to redshifts $z\approx2$~--~3  if it roughly follows the shape of the cosmic star formation rate density of massive stars \cite{doi:10.1146/annurev-astro-081811-125615,2022ApJ...931...17V}.
In addition, the detection catalogue of the first three LIGO-Virgo observing runs contains two BNSs and four NSBH binaries \cite{2023PhRvX..13a1048A}. The inferred merger rates in the local Universe are

\begin{align}
\mathscr{R}_{\rm BNS}&=10\,\mathrm{-}\,1700\,\rm Gpc^{-3}\,yr^{-1}\quad(90\%\,\rm C.I.),\label{eq:RBNS}\\
\mathscr{R}_{\rm NSBH}&=7.8\,\mathrm{-}\,140\,\rm Gpc^{-3}\,yr^{-1}\quad(90\%\,\rm C.I.).
\end{align}

\noindent Due to the small number of detections these rate estimates are quite uncertain.
Furthermore, no redshift evolution could be determined in either cases. For the purpose of this work, we agnostically assume the same redshift scaling as for the BBH mergers. 

Regarding the chirp mass of the mergers, the BBHs follow a clumped distribution with pronounced overdensities around $\mathscr{M}=8.3^{+0.3}_{-0.5}\,\rm M_\odot$ and $27.9^{+1.9}_{-1.8}\,\rm M_\odot$. Meanwhile, the chirp masses of the six events involving a neutron star range from $\mathscr{M}=1.188^{+0.004}_{-0.002}\,\rm M_\odot$ (GW170817 \cite{2017PhRvL.119p1101A}) to $3.7\pm0.2\,\rm M_\odot$ (GW190917 \cite{2021arXiv210801045T}). The chirp mass of the carriers affects our results in two ways. On the one hand, a higher chirp mass increases the SNR of the carrier which overall benefits the identification of a modulating background source. On the other hand, a higher chirp mass shortens the merger time and the time a binary spends in the frequency bandwidth of a detector [cf., equation~\eqref{eq:T}]. This impedes the identification of a modulating background source in a more significant way. To demonstrate this effect, we adopt $\mathscr{M}=27.9\,\rm M_\odot$ from the first directly detected BBH merger (GW150914\cite{GW150914}) as a fiducial value for the chirp mass of a rather massive equal-mass BBH, and assume the chirp mass $\mathscr{M}=1.188\,\rm M_\odot$ of an equal-mass BNS as our standard choice for binaries with a neutron star component.

\subsection*{Proposed deci-Hz detectors}\label{sec:detector-noise}
A GW detector that is sensitive to frequencies around $\gtrsim\mathscr{O}(10^{-2})\,\rm Hz$ is ideal for our type of analysis. In this frequency range, the compact binaries  that we consider in this work spend a time $T\lesssim f/\dot{f}$ up to several years before they merge in the bandwidth of LIGO-Virgo. Monitoring carrier binaries over several years allows us to detect frequencies of a modulating background source as low as $f_{\rm mod}\gtrsim1/T\sim\mathscr{O}(10^{-9})\,\rm Hz$, which would be a probe of the most massive black hole binaries in the Universe. For this reason, we focus on existing detector proposals in the deci-Hz regime. For concreteness, we consider the space-based instruments DECIGO \cite{2011CQGra..28i4011K,2021PTEP.2021eA105K} and BBO \cite{PhysRevD.72.083005,2006CQGra..23.4887H,2006CQGra..23..C01H}, but stress that our analysis could be applied to any detector which is sensitive to those frequencies. We note that DECIGO is actively being developed, while we are not aware of any research efforts towards the realisation of BBO. 

Both DECIGO and BBO  consist of multiple (up to four) triangular constellations of spacecraft in heliocentric orbits, where the constellations are equally spaced around the ecliptic. DECIGO and BBO are designed to detect GWs at frequencies between 0.1~Hz and 10~Hz, such that they could provide sensitivity in a frequency range between the sensitive band of LISA and ground-based detectors such as LIGO-Virgo. Both DECIGO and BBO are laser-interferometric detectors where the interferometer arms are rendered by beams of laser light that travel between the spacecraft.
In general, the sensitivity of interferometers decreases above a frequency inversely proportional to the arm length, and therefore both DECIGO and BBO will employ arms shorter than those of LISA. DECIGO plans to use interferometer arm lengths $\sim10^3\,\rm km$; BBO's arms are intended to have a length of $\sim5\times10^4\,\rm km$ (cf., LISA which has a design with arm length $\sim2.5\times10^6\,\rm km$)\cite{2017arXiv170200786A,2023LRR....26....2A,2006CQGra..23.4887H}. While BBO has arms that are longer than those of DECIGO by a factor $\sim10$, DECIGO is designed with arms that contain Fabry-Perot cavities with a finesse $\sim10$, and the instruments therefore have almost equal strain sensitivities. The optical frequency response of both instruments is comparable as well: the sensitivity to GWs falls off linearly at frequencies above the pole frequency of the Fabry-Perot cavities in the arms of DECIGO, and similary above the light-crossing frequency of the arms of BBO. The design of DECIGO envisions the instrument to be limited by quantum noise across the entire band, specifically quantum radiation pressure noise and quantum shot noise. Extensive mitigation of classical low-frequency acceleration noise will be required to achieve this condition. Quantum radiation pressure noise falls off as $f^{-2}$, and quantum shot noise is constant at all frequencies; the crossover point of these two noise contributions is determined by the optical power circulating in the arms. BBO is expected to be limited at low frequency by acceleration noise of the test masses due to a variety of sources, and at high frequencies by quantum shot noise. 

For both BBO and DECIGO, the triangular design of the instrument provides two independent interferometric measurements of incident GWs, which can be equivalently described as measurements from two independent L-shaped Michelson interferometers \cite{2017arXiv170200786A,2019CQGra..36j5011R,2011PhRvD..83d4011Y,2017PhRvD..95j9901Y}. Given the proposed experimental parameters, the quantum noise and optical response may be modelled, and the sensitivity of of each independent L-shaped interferometer can thus be described as an effective noise power spectral density $S_n$. Here, we use the analytical expressions \cite{2011PhRvD..83d4011Y,2017PhRvD..95j9901Y} for DECIGO

\begin{align}
    S_n^{\rm DECIGO}(f) =\,&7.05\times10^{-48}\left[1+\left(\frac{f}{f_p}\right)^2\right]{\,\rm Hz^{-1}}\nonumber\\
    &+4.8\times10^{-51}\left(\frac{f}{1\,\rm Hz}\right)^{-4}\frac{1}{1+\left(\frac{f}{f_p}\right)^2}{\,\rm Hz^{-1}}\nonumber\\
    &+5.33\times10^{-52}\left(\frac{f}{1\,\rm Hz}\right)^{-4}{\,\rm Hz^{-1}},\label{eq:DECIGO}
\end{align}

\noindent where $f_p=7.35\,\rm Hz$ and for BBO

\begin{align}
    S_n^{\rm BBO}(f)=\,&2.00\times10^{-49}\left(\frac{f}{1\,\rm Hz}\right)^2{\,\rm Hz^{-1}}\nonumber\\
    &+4.58\times10^{-49}{\,\rm Hz^{-1}}\nonumber\\
    &+1.26\times10^{-51}\left(\frac{f}{1\,\rm Hz}\right)^{-4}{\,\rm Hz^{-1}}.\label{eq:BBO}
\end{align}

\noindent The sensitivity further depends on the implementation of the instruments; the achievable SNR increases as $\sqrt{N_\mathrm{IFO}}$, where $N_\mathrm{IFO}$ is the number of independent L-shaped Michelson interferometers that the detector furnishes. In this paper, for both DECIGO and BBO, we take a sensitivity corresponding to one triangular constellation, such that $N_\mathrm{IFO}=2$, i.e., we divide Eqs.~\eqref{eq:DECIGO} and~\eqref{eq:BBO} by $\sqrt{2}$. Furthermore, we assume a nominal mission lifetime for both instruments of $T=10\,\rm yr$.
In assessing the sensitvity of DECIGO/BBO, we also consider the angular response factor $Q$ (cf., equation~\eqref{eq:unmodulated-waveform}). Averaging over the antenna pattern for sky positions ($\vartheta,\varphi$) of an L-shaped Michelson and all possible binary orbital inclinations ($\iota$) gives \cite{MaggioreMichele2008GWV1}

\begin{equation}
    \sqrt{\langle|Q(\vartheta,\varphi;\iota)|^2\rangle}= \frac{2}{5},
\end{equation}

\noindent which we will use for $Q$ in this work.

\subsection*{Background source population}\label{sec:background-population}

\begin{figure}
    \centering
    \includegraphics[width=0.46\linewidth]{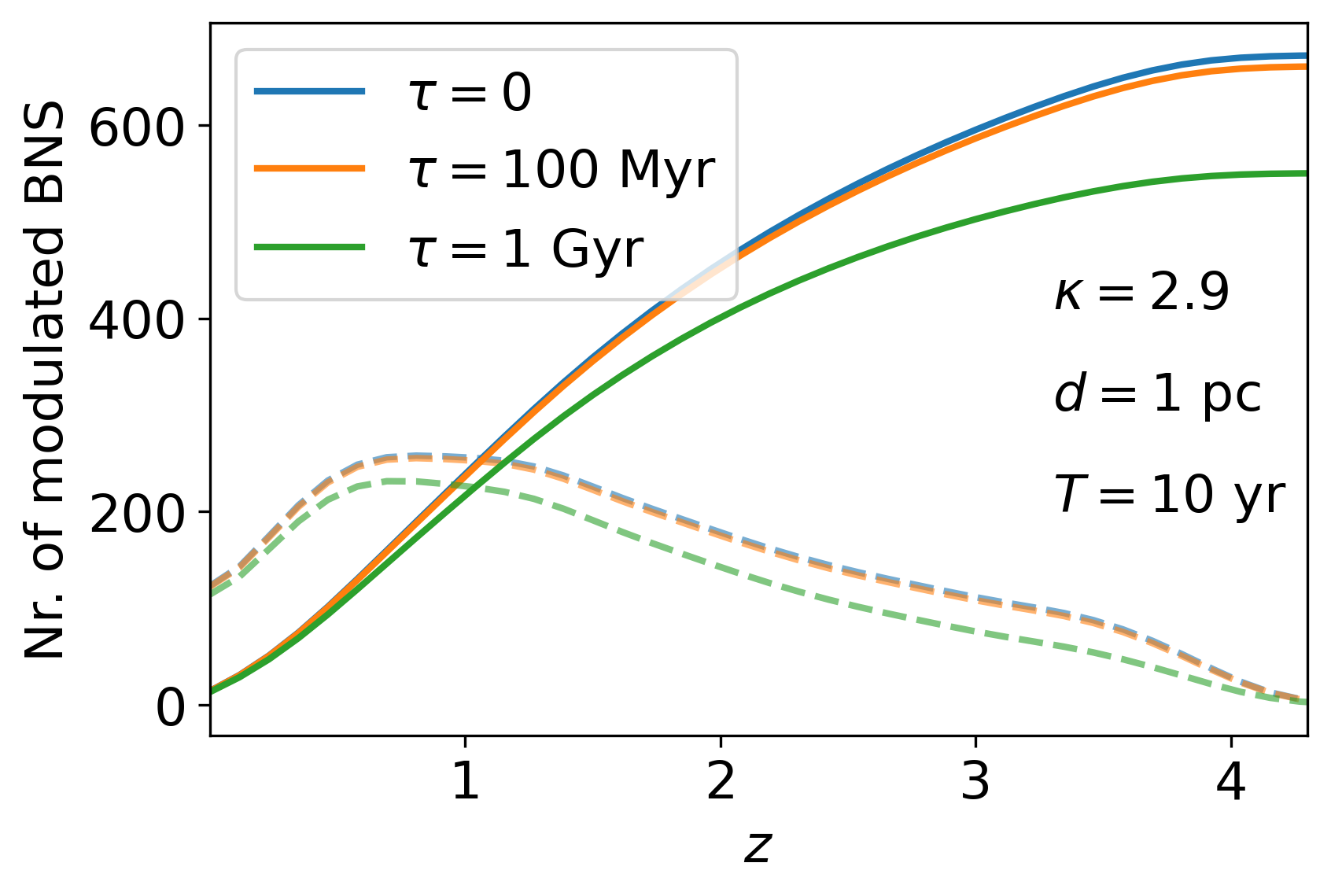}
     \includegraphics[width=0.48\linewidth]{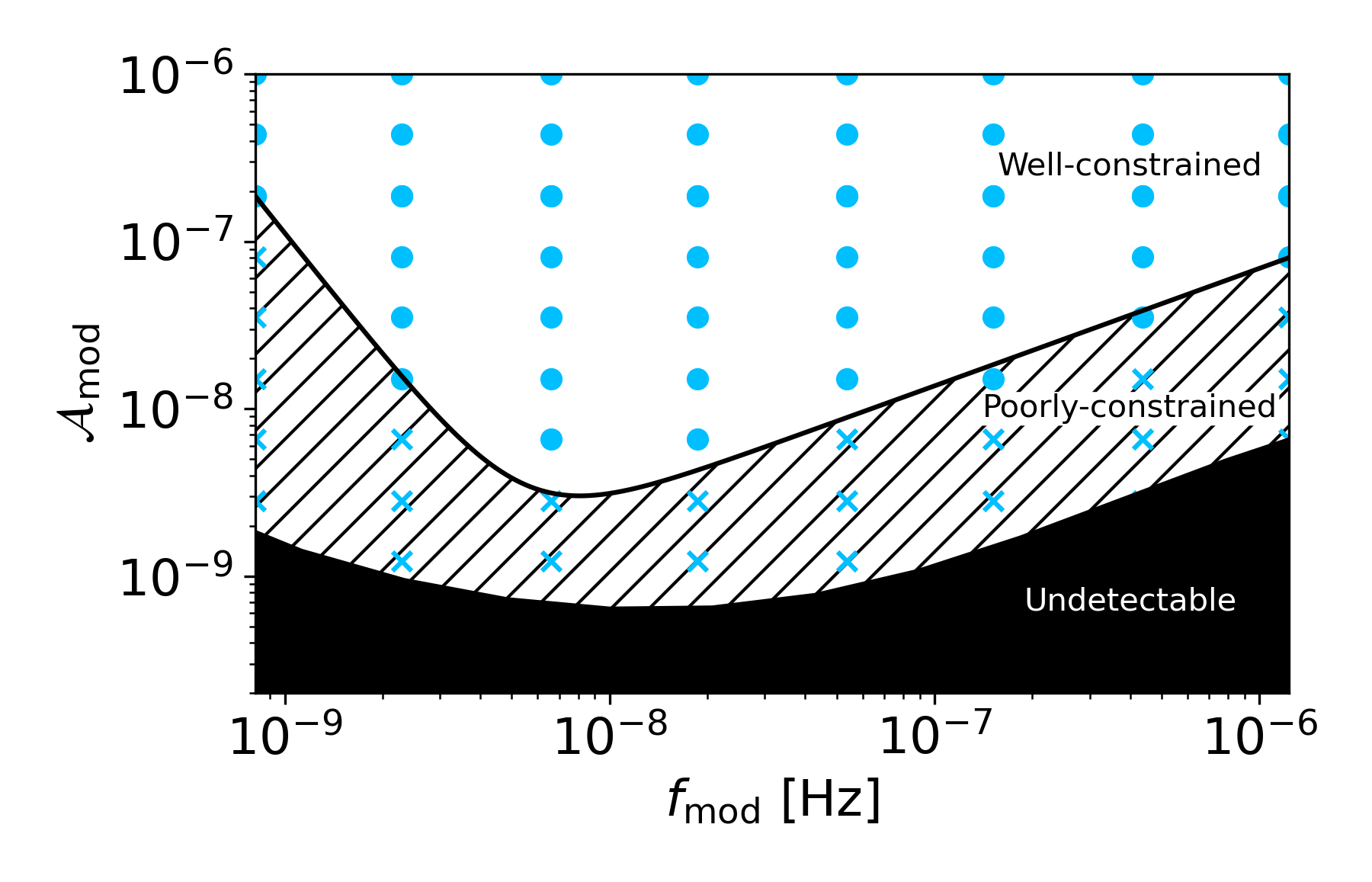}
    \caption{{\textbf{Additional tests.} \textit{Left panel:} We show how the differential (dashed lines) and cumulative (solid lines) distribution of modulated BNSs varies as a function of a constant delay time $\tau$ between halo mergers and SMBHB mergers. \textit{Right panel:} Each marker corresponds to a full Bayesian parameter estimation with an MCMC, e.g., as in Fig.~\ref{fig:MCMC}, of a modulated BNS at our exemplary redshift $z=0.84$. The axes define the amplitude and frequency of the injected modulation. Dots and crosses indicate whether the modulation could be resolved with the MCMC or not, respectively. We define a modulation to be {well-constrained} if the one-sigma width of the posterior distribution of $\mathscr{A}_{\rm mod}$ is smaller than the injected value, e.g., as in Fig.~\ref{fig:MCMC}. A sensitivity curve is constructed (solid {black} line) by {fitting} a smoothly broken power law of the form {$\mathscr{A}_{\rm mod}(f_{\rm mod})=a[(f_{\rm mod}/f_0)^\alpha+(f_{\rm mod}/f_0)^\beta]$} to the boundary of {well-constrained} and {poorly-constrained} modulations {(hatched region)} \cite{2015CQGra..32e5004M}. This accommodates different slopes of the sensitivity below and above the frequency at peak sensitivity and is used in Fig.~\ref{fig:Amod-fmod}.} {While the hatched region demarcates the region where any present modulation would be poorly resolved, the black shaded region defines the parameter space where no modulation could be confidently established at all. To construct this region we have run an additional MCMC with the null hypothesis of no modulation being injected to the data. Inside the black region the modulation was found small enough to be consistent with the null hypothesis whereas outside we conclude the modulation to be distinguishable from zero.}}
    \label{fig:delays}
\end{figure}

The background sources considered in this work consist in individual SMBHBs, which can span over several decades in mass, mass ratio and orbital separation. The SMBHB merger rate is highly uncertain, and constraining it is one of the major scientific goals of proposed GW observatories such as LISA \cite{2017arXiv170200786A,2023LRR....26....2A}, TianQin \cite{2019PhRvD.100d3003W}, pulsar timing arrays \cite{2013CQGra..30v4010M,2015MNRAS.453.2576L,2015aska.confE..37J,2020ApJ...905L..34A,2023arXiv230103608A}, and other electromagnetic searches \cite{2013ApJ...777...44J,2022LRR....25....3B}. For the purposes of this work, we broadly follow the procedure detailed in \cite{2022MNRAS.511.6143Z}, and link the SMBHB merger rate to the well-established halo merger rate, based on the two Millenium simulations \cite{2010MNRAS.406.2267F}:

\begin{align}
    \label{eq:millennium}
    \frac{\d ^2 H}{\d \xi \d z} = B_1 \left( \frac{M_{\rm{halo}}}{10^{12} \, \rm{M}_{\odot}} \right)^{b_1} \xi ^{b_2} \exp \left[\left( \frac{\xi}{B_2} \right)^{b_3}\right] (1+z)^{b_4},
\end{align}

\noindent where $H$ is the total number of mergers that a halo of mass $M_{\rm{halo}}$ experiences over cosmic time, $\xi\leq1$ is the halo merger mass ratio and the best-fit parameters are given by $[B_1,B_2,b_1,b_2,b_3,b_4] = [0.0104,9.72 \times 9.72,0.133,-1.995,0.263,0.0993]$. The SMBHB merger rate $\dot{N}_{\bullet \bullet}$ can be obtained by multiplying the halo merger rate with the SMBH mass function:

\begin{align}
    \label{eq:generalSMBHBrate}
    \frac{\d^{3} \dot{N}_{\bullet \bullet}}{\d M_{\bullet} \d \xi  \d z} = {P}(M_{\rm{halo}},z) \frac{4 \pi D^2_{\rm{com}}(z)}{(1+z)^3} \frac{\d n_{\bullet}}{\d M_{\bullet}} \frac{\d ^2 H}{\d \xi \d z}(\xi,z_{\rm{del}}),
\end{align}

\noindent where we must supply an occupation fraction $P$, a relation between SMBH and halo masses and a delay prescription between the nominal halo merger and the actual SMBHB merger\footnote{Note that $c=1$.}. Considering that the most easily detectable modulations will be naturally produced by heavy, low-redshift binaries we can tailor equation~\eqref{eq:generalSMBHBrate} to SMBH masses of $M_{\bullet} > 10^8$ M$_{\odot}$ within moderately low redshifts $z< 4$. Firstly, we adopt the SMBH mass function reported in Ref.~\cite{2013MNRAS.428..421S} \footnote{We interpolate the reported curves in Fig. 3 for redshifts of $0 < z < 4$ and extrapolate the curves for $M_{\bullet}> 10^{10}$ M$_{\odot}$.} which we denote as

\begin{align}
    \phi\equiv\frac{\d n_{\bullet}}{\d \log_{10} M_{\bullet}},
\end{align}

\noindent where here, $\phi$ is the symbol customarily used to denote the SMBH mass function in units of Mpc$^{-3}$ yr$^{-1}$ dex$^{-1}$. Secondly, we adopt a simple relation between halo and SMBH mass from \cite{2009MNRAS.394.1109C}:

\begin{align}
     M_{\bullet}&=\left[\frac{M_{\rm{halo}}(1+z)}{2\times10^7\,\rm M_\odot}\right]^{3/2}\,\rm M_\odot,\label{eq:M_bullet}
\end{align}

\noindent which is broadly consistent with both theoretical and observational constraints \cite{2021MNRAS.507.4274M,2023MNRAS.tmp.1567B}. Given the selection effects towards higher masses, we can safely assume an occupation fraction $P\approx1$ as it is strongly suggested by simulations \cite{2018ricarte} and observations of AGN \cite{2019MNRAS.487..275G}. Finally, the time delays between the halo merger and the eventual SMBHB merger are expected to be of the order $10^8$~--~$10^9\,\rm yr$ for the SMBHBs we are considering \cite{2015ApJ...810...49V,2017ApJ...840...31D}. As seen in Fig.~\ref{fig:delays}, we have tested the effect of introducing constant time delays of up to 1 Gyr by shifting the merger redshift to $z \to z - \Delta z(\tau)$, where $\tau$ is the time delay and we assume a standard cosmology. We find that our results do not change appreciably, decreasing by at most $\sim 20 \%$ and only at redshifts higher than $\sim 3$. {This result runs counter to the common expectation that introducing delays significantly reduces the number of detected massive BH mergers for GW detectors \cite{2016PhRvD..93b4003K,2024arXiv240207571C}. In our work, it can be attributed to two factors. Firstly, the halo merger rate reported in the Millennium simulation only weakly depends on redshift, scaling as $\sim(1+z)^{0.1}$. Secondly, in our setup shifting SMBHB mergers to lower redshift automatically increases the SNR of the detected stellar-mass compact object binaries while simultaneously reducing their cosmological redshift. This has the consequence to greatly expand the range of detectable modulations in both frequency and amplitude, counteracting the diminished number of SMBHB mergers. The fact that these effects seem to cancel out each other is somewhat of a coincidence, which we attribute to the explicit form of equation~\eqref{eq:millennium}.}

\subsection*{Rate estimate}\label{sec:rates-methods}
Given a population of both carriers and background sources, we can estimate the mean number of compact binaries that could exhibit a detectable modulation. In order to achieve this, we must select all SMBHBs that are emitting loud GWs in the appropriate frequency range for the sensitivity curves detailed above (cf., Fig.~\ref{fig:Amod-fmod}). The first step is to distribute the SMBHB over several frequency bins, by replacing the time derivative with an integration over $f_{\rm{mod}}$:

\begin{align}
    N_{\bullet \bullet} = \int \dot{N}_{\bullet \bullet} {\rm d}t = \frac{{\rm{d}} N_{\bullet \bullet}}{{\rm d} f_{\rm{mod}}} {\dot{f}_{\rm{mod}}} {\rm d}t =\int \frac{{\rm{d}} N_{\bullet \bullet}}{{\rm d} f_{\rm{mod}}} \,\d f_{\rm{mod}}
\end{align}

\noindent Note that, since we are considering SMBHBs in the GW-driven regime, we can simply use equation~\eqref{eq:f(t)} to describe the frequency evolution. Then, we can integrate the differential contributions over frequency and mass ratio, using a Heaviside function $\Theta$ to only select SMBHB that would produce detectable modulations, as defined by a certain confidence threshold $\sigma_{\rm{det}}$:

\begin{align}
   \frac{\d ^{2} N_{\bullet \bullet}^{\rm{mod}}}{\d M_{\bullet} \d z } = \int \int  \frac{\d^{4} N_{\bullet \bullet}}{\d M_{\bullet} \d \xi \d f_{\rm{mod}} \d z} \Theta \left(\sigma_{\rm{det}}- \frac{\Delta \mathscr{A}_{\rm{mod}}}{\mathscr{A}_{\rm{mod}}}\right) \,\d f_{\rm{mod}} \,\d \xi.
\end{align}

\noindent The Heaviside function depends on the sensitivity curves detailed above (cf., Fig.~\ref{fig:Amod-fmod}):

\begin{align}
    \Theta =\Theta(z,\mathscr{A}_{\rm mod},f_{\rm mod},\mathscr{M},t_c,\phi_c,\phi_{\rm mod}),
\end{align}

\noindent where we initially choose a threshold of $\Delta \mathscr{A}_{\rm{mod}} = \mathscr{A}_{\rm{mod}}$, i.e., $\sigma_{\rm{det}}=1$, to represent a SMBHB modulation the parameters of which can be well constrained. {Note that, typically, constraining the parameters of a GW to $1\sigma$ does not constitute a sufficient threshold to avoid the possibility of false signals. In our case however, we are considering modulations that affect GWs signals with SNR of order few $10^2$, and it is not immediately clear whether the same intuition applies, nor if the posterior distributions of the parameters are Gaussian. To test this, we have run additional MCMC tests in which the null hypothesis, i.e., the absence of a signal, is tested against several realisations of Gaussian noise. We observe that the MCMC walkers freely sample regions in which the trial amplitude is very low. However, they are strongly discouraged to sample large values of $\mathscr{A}_{\rm{mod}}$, which are inconsistent with the absence of a signal. The posterior probability for the recovered modulation amplitude $\mathscr{A}_{\rm mod}$ sharply decreases above a certain threshold $\mathscr{A}_{\rm NH}$ associated to the null-hypothesis:
\begin{align}
    \mathcal{P}(\mathscr{A}_{\rm mod}) \sim \begin{cases} {\text{const.}} & {\text{for }} \mathscr{A}_{\rm mod} < \mathscr{A}_{\rm NH}(f_{\rm mod})\\
     0 & {\text{for }} \mathscr{A}_{\rm mod} > \mathscr{A}_{\rm NH}(f_{\rm mod}).
    \end{cases}
\end{align}
More precisely, the boundary is a half-Gaussian, for which the inflection point defines the $1\sigma_{\rm NH}$ confidence threshold between the allowed and disallowed regions. We find that the boundary at which the null hypothesis can be confidently excluded always lies about one order of magnitude below our choice of sensitivity curve, especially at lower modulation frequencies, where the majority of detections take place (see Fig. \ref{fig:delays}). In this sense, our choice of using a one sigma boundary simply refers to the desired accuracy in the recovered modulation parameters, though it is likely that a more complex noise model would affect this conclusion. Therefore, the one sigma detection criterion of $\mathscr{A}_{\rm mod}$ does only define how accurate we could determine its value when a modulation is present and does not imply a large false positive probability at which we would wrongfully conclude the existence of a modulation. We have additionally tested how the number of detection scales as a function of the desired posterior widths, finding that:
\begin{align}
    N_{\rm det}(\sigma_{\rm det}) \propto  \sigma_{\rm det}^{-1.2}.
\end{align}
This means that a significant fraction of the detections will result in better constrained posteriors.}

The next step is to distribute the total amount of compact object mergers into their host galaxies and across cosmic time. To enable this calculation, we require to know the differential compact object binary rate per halo. The most natural assumption is that the latter must be proportional to the stellar mass of the galaxy that occupies a halo of a given size. Interestingly, for the high-mass galaxies likely to host heavy SMBHs, the stellar mass-halo mass relation has a turnover point at a value of $\sim 1/100$, varying by approximately one order of magnitude over the large range of $10^{11}$ M$_\odot$ $< M_{\rm{halo}} <$ $10^{14}$ M$_\odot$. For the purposes of this work, we use a fit of the stellar mass to halo mass ratio $\mathscr{F}^{\star}$ \cite{2018MNRAS.477.1822M,2020A&A...634A.135G}:

\begin{align}
    \mathscr{F}^{\star}(M_{\rm{halo}}) &= 2D_1(1+z)^{\delta_1}\left[\left(\frac{M_{\rm{Halo}}}{D_2} \right)^{-\beta} + \left(\frac{M_{\rm{Halo}}}{D_2} \right)^{\gamma} \right]^{-1},\\
    D_2&= 10^{\delta_2 + z\delta_3}\, {\rm M}_{\odot},\\
    \beta &= z\delta_4 + \delta_5,\\
    \gamma &= \delta_6(1+z)^{\delta_7},
\end{align}

\noindent where the best fit parameters are $[D_1,\delta_1,\delta_2,\delta_3,\delta_4,\delta_5,\delta_6,\delta_7] = [0.046,-0.38,11.79,0.20,0.043,0.96,0.709,-0.18]$.
Thus, the number of compact object binaries is distributed as:

\begin{align}
    \frac{ \d \mathscr{R}_{\rm{COM}}}{\d M_{\rm{halo}}} = \frac{\mathscr{R}_{\rm{COM}}}{\mathscr{N}}\mathscr{F}^{\star}(M_{\rm{halo}}) M_{\rm{halo}}\frac{\d n_{\rm halo}}{\d M_{\rm{halo}}},
\end{align}

\noindent where the normalisation is given by an integral over the total stellar mass contained in all halos in the Universe:

\begin{align}
   \mathscr{N} = \int \mathscr{F}^{\star}(M_{\rm{halo}}) M_{\rm{halo}}\frac{\d n_{\rm halo}}{\d M_{\rm{halo}}} \,\d M_{\rm{halo}},
\end{align}

\noindent and we obtain the halo mass function by inverting equation~\eqref{eq:M_bullet} and adjusting the differentials accordingly. Note that relaxing the assumption of a constant occupation fraction does not affect the normalisation significantly, because a large fraction of the total mass budget is composed by massive halos above $\sim$10$^{12}$ M$_\odot$.

To estimate the probability of a compact object merger taking place in a galaxy that simultaneously hosts a modulating SMBHB, we suppress the former's rate by a further factor representing the number of halos that contain such an SMBHB with respect to the total number of halos, in a given redshift shell with volume $\d V_z$:

\begin{align}
    \label{eq:interesting_sources}
    \frac{ \d^2  \mathscr{R}_{\rm{COM}}^{\rm{mod}}}{\d M_{\rm{halo}} \d z} = \frac{ \d ^2 \mathscr{R}_{\rm{COM}}}{\d M_{\rm{halo}} \d z} \frac{\d ^{2} N_{\bullet \bullet}^{\rm{mod}}}{\d M_{\rm{halo}} \d z } \left(\frac{\d n_{\rm{halo}} }{\d M_{\rm{halo}}} \frac{\d V_z}{\d z} \right)^{-1},
\end{align}

\noindent where once again we make use of the halo mass-SMBH mass relation and adjust the differentials accordingly. Note that here the compact object merger rate must be transformed from the local source frame to the observer frame, by dividing the observed redshift dependence by a factor $(1+z)$ and expliciting the differential $\d / \d z$.

Finally, the total number of compact object mergers with a detectable modulation $N_{\rm{COM}}^{\rm{mod}}$ can be obtained by integrating and multiplying by the total observation time:

\begin{align}
  N_{\rm{COM}}^{\rm{mod}}  = T \int \frac{ \d^2  \mathscr{R}^{\rm{mod}}}{\d M_{\rm{halo}} \d z} \d M_{\rm{halo}} \,\d z \equiv N_{\rm{det}},
\end{align}

\noindent where, equivalently, $N_{\rm{det}}$ is the number of indirectly detected SMBHB. Note that in reality the number $N_{\rm{COM}}^{\rm{mod}}$ is actually the expectation value of a process that involves sampling both compact object binaries and SMBHB from a large pool of halos, analogous to a binomial distribution\footnote{Excluding the unlikely possibility of having two separate compact object binaries merging simultaneously with an SMBHB in the same galaxy.}. Therefore, it carries an intrinsic uncertainty approximately proportional to its square root. 

\section*{Data Availability}
The data underlying this article will be shared on reasonable request to the corresponding author.

\section*{Code Availability}
The codes underlying this article will be shared on reasonable request to the corresponding author.

\section*{Acknowledgements}
J.S. thanks the Institute for Computational Science at the University of Zurich for the hospitality and acknowledges funding from the Netherlands Organisation for Scientific Research (NWO), as part of the Vidi research program BinWaves (project number 639.042.728, PI: de Mink). L.Z. and L.M. acknowledge support from the Swiss National Science Foundation under the Grant 200020\_192092. L.Z. also acknowledges support from  ERC Starting Grant No. 121817–BlackHoleMergs. S.V. acknowledges funding from the Leverhulme Trust in the UK under research grant RPG-2019-022. F.A. acknowledges support from a Rutherford fellowship (ST/P00492X/1) from the Science and Technology Facilities Council. We thank the anonymous referees for useful suggestions and for reviewing this manuscript.

\section*{Author contributions statement}
J.S. and L.Z. conceived the study, performed the analytical and numerical analysis, and wrote the manuscript. S.V. provided the description of the detectors. All authors contributed to the interpretation of the results and reviewed and edited the manuscript.

\section*{Competing interests}
The authors declare no competing interests.

\bibliography{sample}

\begin{thebibliography}{100}
\urlstyle{rm}
\expandafter\ifx\csname url\endcsname\relax
  \def\url#1{\texttt{#1}}\fi
\expandafter\ifx\csname urlprefix\endcsname\relax\def\urlprefix{URL }\fi
\expandafter\ifx\csname doiprefix\endcsname\relax\def\doiprefix{DOI: }\fi
\providecommand{\bibinfo}[2]{#2}
\providecommand{\eprint}[2][]{\url{#2}}

\bibitem{2020CQGra..37u5011A}
\bibinfo{author}{{Arca Sedda}, M.} \emph{et~al.}
\newblock \bibinfo{journal}{\bibinfo{title}{{The missing link in
  gravitational-wave astronomy: discoveries waiting in the decihertz range}}}.
\newblock {\emph{\JournalTitle{Classical and Quantum Gravity}}}
  \textbf{\bibinfo{volume}{37}}, \bibinfo{pages}{215011},
  \doiprefix\url{10.1088/1361-6382/abb5c1} (\bibinfo{year}{2020}).
\newblock \eprint{1908.11375}.

\bibitem{2011PhRvD..83d4011Y}
\bibinfo{author}{{Yagi}, K.} \& \bibinfo{author}{{Seto}, N.}
\newblock \bibinfo{journal}{\bibinfo{title}{{Detector configuration of
  DECIGO/BBO and identification of cosmological neutron-star binaries}}}.
\newblock {\emph{\JournalTitle{Physical Review D}}}
  \textbf{\bibinfo{volume}{83}}, \bibinfo{pages}{044011},
  \doiprefix\url{10.1103/PhysRevD.83.044011} (\bibinfo{year}{2011}).
\newblock \eprint{1101.3940}.

\bibitem{2017PhRvD..95j9901Y}
\bibinfo{author}{{Yagi}, K.} \& \bibinfo{author}{{Seto}, N.}
\newblock \bibinfo{journal}{\bibinfo{title}{{Erratum: Detector configuration of
  DECIGO/BBO and identification of cosmological neutron-star binaries [Phys.
  Rev. D 83, 044011 (2011)]}}}.
\newblock {\emph{\JournalTitle{Physical Review D}}}
  \textbf{\bibinfo{volume}{95}}, \bibinfo{pages}{109901},
  \doiprefix\url{10.1103/PhysRevD.95.109901} (\bibinfo{year}{2017}).

\bibitem{2016PhRvL.116x1102A}
\bibinfo{author}{{Abbott}, B.~P.} \emph{et~al.}
\newblock \bibinfo{journal}{\bibinfo{title}{{Properties of the Binary Black
  Hole Merger GW150914}}}.
\newblock {\emph{\JournalTitle{Physical Review Letters}}}
  \textbf{\bibinfo{volume}{116}}, \bibinfo{pages}{241102},
  \doiprefix\url{10.1103/PhysRevLett.116.241102} (\bibinfo{year}{2016}).
\newblock \eprint{1602.03840}.

\bibitem{2017PhRvL.119p1101A}
\bibinfo{author}{{Abbott}, B.~P.} \emph{et~al.}
\newblock \bibinfo{journal}{\bibinfo{title}{{GW170817: Observation of
  Gravitational Waves from a Binary Neutron Star Inspiral}}}.
\newblock {\emph{\JournalTitle{Physical Review Letters}}}
  \textbf{\bibinfo{volume}{119}}, \bibinfo{pages}{161101},
  \doiprefix\url{10.1103/PhysRevLett.119.161101} (\bibinfo{year}{2017}).
\newblock \eprint{1710.05832}.

\bibitem{2015CQGra..32e5004M}
\bibinfo{author}{{Moore}, C.~J.}, \bibinfo{author}{{Taylor}, S.~R.} \&
  \bibinfo{author}{{Gair}, J.~R.}
\newblock \bibinfo{journal}{\bibinfo{title}{{Estimating the sensitivity of
  pulsar timing arrays}}}.
\newblock {\emph{\JournalTitle{Classical and Quantum Gravity}}}
  \textbf{\bibinfo{volume}{32}}, \bibinfo{pages}{055004},
  \doiprefix\url{10.1088/0264-9381/32/5/055004} (\bibinfo{year}{2015}).
\newblock \eprint{1406.5199}.

\bibitem{2019PhRvD.100j4028H}
\bibinfo{author}{{Hazboun}, J.~S.}, \bibinfo{author}{{Romano}, J.~D.} \&
  \bibinfo{author}{{Smith}, T.~L.}
\newblock \bibinfo{journal}{\bibinfo{title}{{Realistic sensitivity curves for
  pulsar timing arrays}}}.
\newblock {\emph{\JournalTitle{Physical Review D}}}
  \textbf{\bibinfo{volume}{100}}, \bibinfo{pages}{104028},
  \doiprefix\url{10.1103/PhysRevD.100.104028} (\bibinfo{year}{2019}).
\newblock \eprint{1907.04341}.

\bibitem{2022PhRvD.105d4005B}
\bibinfo{author}{{Bustamante-Rosell}, M.~J.}, \bibinfo{author}{{Meyers}, J.},
  \bibinfo{author}{{Pearson}, N.}, \bibinfo{author}{{Trendafilova}, C.} \&
  \bibinfo{author}{{Zimmerman}, A.}
\newblock \bibinfo{journal}{\bibinfo{title}{{Gravitational wave timing
  array}}}.
\newblock {\emph{\JournalTitle{Physical Review D}}}
  \textbf{\bibinfo{volume}{105}}, \bibinfo{pages}{044005},
  \doiprefix\url{10.1103/PhysRevD.105.044005} (\bibinfo{year}{2022}).
\newblock \eprint{2107.02788}.

\bibitem{LIGOcurve}
\bibinfo{author}{{Barsotti}, L.}, \bibinfo{author}{{Fritschel}, P.},
  \bibinfo{author}{{Evans}, M.} \& \bibinfo{author}{{Gras}, S.}
\newblock \bibinfo{title}{Updated advanced ligo sensitivity design curve
  (ligo-t1800044-v5)} (\bibinfo{year}{2018}).

\bibitem{2008arXiv0810.0604H}
\bibinfo{author}{{Hild}, S.}, \bibinfo{author}{{Chelkowski}, S.} \&
  \bibinfo{author}{{Freise}, A.}
\newblock \bibinfo{journal}{\bibinfo{title}{{Pushing towards the ET sensitivity
  using 'conventional' technology}}}.
\newblock {\emph{\JournalTitle{arXiv e-prints}}}
  \bibinfo{pages}{arXiv:0810.0604}, \doiprefix\url{10.48550/arXiv.0810.0604}
  (\bibinfo{year}{2008}).
\newblock \eprint{0810.0604}.

\bibitem{ETcurve}
\bibinfo{author}{{Punturo}, M.}
\newblock \bibinfo{title}{Et sensitivities page} (\bibinfo{year}{2021}).

\bibitem{CEcurve}
\bibinfo{author}{{The CE Consortium}}.
\newblock \bibinfo{title}{Astrophysical sensitivity} (\bibinfo{year}{2020}).

\bibitem{2009LRR....12....2S}
\bibinfo{author}{{Sathyaprakash}, B.~S.} \& \bibinfo{author}{{Schutz}, B.~F.}
\newblock \bibinfo{journal}{\bibinfo{title}{{Physics, Astrophysics and
  Cosmology with Gravitational Waves}}}.
\newblock {\emph{\JournalTitle{Living Reviews in Relativity}}}
  \textbf{\bibinfo{volume}{12}}, \bibinfo{pages}{2},
  \doiprefix\url{10.12942/lrr-2009-2} (\bibinfo{year}{2009}).
\newblock \eprint{0903.0338}.

\bibitem{2015CQGra..32a5014M}
\bibinfo{author}{{Moore}, C.~J.}, \bibinfo{author}{{Cole}, R.~H.} \&
  \bibinfo{author}{{Berry}, C.~P.~L.}
\newblock \bibinfo{journal}{\bibinfo{title}{{Gravitational-wave sensitivity
  curves}}}.
\newblock {\emph{\JournalTitle{Classical and Quantum Gravity}}}
  \textbf{\bibinfo{volume}{32}}, \bibinfo{pages}{015014},
  \doiprefix\url{10.1088/0264-9381/32/1/015014} (\bibinfo{year}{2015}).
\newblock \eprint{1408.0740}.

\bibitem{Bartos_2017}
\bibinfo{author}{Bartos, I.}, \bibinfo{author}{Kocsis, B.},
  \bibinfo{author}{Haiman, Z.} \& \bibinfo{author}{Márka, S.}
\newblock \bibinfo{journal}{\bibinfo{title}{Rapid and bright stellar-mass
  binary black hole mergers in active galactic nuclei}}.
\newblock {\emph{\JournalTitle{The Astrophysical Journal}}}
  \textbf{\bibinfo{volume}{835}}, \bibinfo{pages}{165},
  \doiprefix\url{10.3847/1538-4357/835/2/165} (\bibinfo{year}{2017}).

\bibitem{2017MNRAS.464..946S}
\bibinfo{author}{{Stone}, N.~C.}, \bibinfo{author}{{Metzger}, B.~D.} \&
  \bibinfo{author}{{Haiman}, Z.}
\newblock \bibinfo{journal}{\bibinfo{title}{{Assisted inspirals of stellar mass
  black holes embedded in AGN discs: solving the `final au problem'}}}.
\newblock {\emph{\JournalTitle{Monthly Notices of the Royal Astronomical
  Society}}} \textbf{\bibinfo{volume}{464}}, \bibinfo{pages}{946--954},
  \doiprefix\url{10.1093/mnras/stw2260} (\bibinfo{year}{2017}).
\newblock \eprint{1602.04226}.

\bibitem{2018ApJ...866...66M}
\bibinfo{author}{{McKernan}, B.} \emph{et~al.}
\newblock \bibinfo{journal}{\bibinfo{title}{{Constraining Stellar-mass Black
  Hole Mergers in AGN Disks Detectable with LIGO}}}.
\newblock {\emph{\JournalTitle{The Astrophysical Journal}}}
  \textbf{\bibinfo{volume}{866}}, \bibinfo{pages}{66},
  \doiprefix\url{10.3847/1538-4357/aadae5} (\bibinfo{year}{2018}).
\newblock \eprint{1702.07818}.

\bibitem{2020A&A...638A.119G}
\bibinfo{author}{{Gr{\"o}bner}, M.}, \bibinfo{author}{{Ishibashi}, W.},
  \bibinfo{author}{{Tiwari}, S.}, \bibinfo{author}{{Haney}, M.} \&
  \bibinfo{author}{{Jetzer}, P.}
\newblock \bibinfo{journal}{\bibinfo{title}{{Binary black hole mergers in AGN
  accretion discs: gravitational wave rate density estimates}}}.
\newblock {\emph{\JournalTitle{Astronomy and Astrophysics}}}
  \textbf{\bibinfo{volume}{638}}, \bibinfo{pages}{A119},
  \doiprefix\url{10.1051/0004-6361/202037681} (\bibinfo{year}{2020}).
\newblock \eprint{2005.03571}.

\bibitem{2020ApJ...896..138Y}
\bibinfo{author}{{Yang}, Y.} \emph{et~al.}
\newblock \bibinfo{journal}{\bibinfo{title}{{Cosmic Evolution of Stellar-mass
  Black Hole Merger Rate in Active Galactic Nuclei}}}.
\newblock {\emph{\JournalTitle{The Astrophysical Journal}}}
  \textbf{\bibinfo{volume}{896}}, \bibinfo{pages}{138},
  \doiprefix\url{10.3847/1538-4357/ab91b4} (\bibinfo{year}{2020}).
\newblock \eprint{2003.08564}.

\bibitem{2020ApJ...898...25T}
\bibinfo{author}{{Tagawa}, H.}, \bibinfo{author}{{Haiman}, Z.} \&
  \bibinfo{author}{{Kocsis}, B.}
\newblock \bibinfo{journal}{\bibinfo{title}{{Formation and Evolution of
  Compact-object Binaries in AGN Disks}}}.
\newblock {\emph{\JournalTitle{The Astrophysical Journal}}}
  \textbf{\bibinfo{volume}{898}}, \bibinfo{pages}{25},
  \doiprefix\url{10.3847/1538-4357/ab9b8c} (\bibinfo{year}{2020}).
\newblock \eprint{1912.08218}.

\bibitem{2022MNRAS.517.5827F}
\bibinfo{author}{{Ford}, K.~E.~S.} \& \bibinfo{author}{{McKernan}, B.}
\newblock \bibinfo{journal}{\bibinfo{title}{{Binary black hole merger rates in
  AGN discs versus nuclear star clusters: loud beats quiet}}}.
\newblock {\emph{\JournalTitle{Monthly Notices of the Royal Astronomical
  Society}}} \textbf{\bibinfo{volume}{517}}, \bibinfo{pages}{5827--5834},
  \doiprefix\url{10.1093/mnras/stac2861} (\bibinfo{year}{2022}).
\newblock \eprint{2109.03212}.

\bibitem{2012ApJ...757...27A}
\bibinfo{author}{{Antonini}, F.} \& \bibinfo{author}{{Perets}, H.~B.}
\newblock \bibinfo{journal}{\bibinfo{title}{{Secular Evolution of Compact
  Binaries near Massive Black Holes: Gravitational Wave Sources and Other
  Exotica}}}.
\newblock {\emph{\JournalTitle{The Astrophysical Journal}}}
  \textbf{\bibinfo{volume}{757}}, \bibinfo{pages}{27},
  \doiprefix\url{10.1088/0004-637X/757/1/27} (\bibinfo{year}{2012}).
\newblock \eprint{1203.2938}.

\bibitem{2017ApJ...846..146P}
\bibinfo{author}{{Petrovich}, C.} \& \bibinfo{author}{{Antonini}, F.}
\newblock \bibinfo{journal}{\bibinfo{title}{{Greatly Enhanced Merger Rates of
  Compact-object Binaries in Non-spherical Nuclear Star Clusters}}}.
\newblock {\emph{\JournalTitle{The Astrophysical Journal}}}
  \textbf{\bibinfo{volume}{846}}, \bibinfo{pages}{146},
  \doiprefix\url{10.3847/1538-4357/aa8628} (\bibinfo{year}{2017}).
\newblock \eprint{1705.05848}.

\bibitem{2018ApJ...856..140H}
\bibinfo{author}{{Hoang}, B.-M.}, \bibinfo{author}{{Naoz}, S.},
  \bibinfo{author}{{Kocsis}, B.}, \bibinfo{author}{{Rasio}, F.~A.} \&
  \bibinfo{author}{{Dosopoulou}, F.}
\newblock \bibinfo{journal}{\bibinfo{title}{{Black Hole Mergers in Galactic
  Nuclei Induced by the Eccentric Kozai-Lidov Effect}}}.
\newblock {\emph{\JournalTitle{The Astrophysical Journal}}}
  \textbf{\bibinfo{volume}{856}}, \bibinfo{pages}{140},
  \doiprefix\url{10.3847/1538-4357/aaafce} (\bibinfo{year}{2018}).
\newblock \eprint{1706.09896}.

\bibitem{2018ApJ...865....2H}
\bibinfo{author}{{Hamers}, A.~S.}, \bibinfo{author}{{Bar-Or}, B.},
  \bibinfo{author}{{Petrovich}, C.} \& \bibinfo{author}{{Antonini}, F.}
\newblock \bibinfo{journal}{\bibinfo{title}{{The Impact of Vector Resonant
  Relaxation on the Evolution of Binaries near a Massive Black Hole:
  Implications for Gravitational-wave Sources}}}.
\newblock {\emph{\JournalTitle{The Astrophysical Journal}}}
  \textbf{\bibinfo{volume}{865}}, \bibinfo{pages}{2},
  \doiprefix\url{10.3847/1538-4357/aadae2} (\bibinfo{year}{2018}).
\newblock \eprint{1805.10313}.

\bibitem{2019MNRAS.488...47F}
\bibinfo{author}{{Fragione}, G.}, \bibinfo{author}{{Grishin}, E.},
  \bibinfo{author}{{Leigh}, N. W.~C.}, \bibinfo{author}{{Perets}, H.~B.} \&
  \bibinfo{author}{{Perna}, R.}
\newblock \bibinfo{journal}{\bibinfo{title}{{Black hole and neutron star
  mergers in galactic nuclei}}}.
\newblock {\emph{\JournalTitle{Monthly Notices of the Royal Astronomical
  Society}}} \textbf{\bibinfo{volume}{488}}, \bibinfo{pages}{47--63},
  \doiprefix\url{10.1093/mnras/stz1651} (\bibinfo{year}{2019}).
\newblock \eprint{1811.10627}.

\bibitem{2019ApJ...878...58S}
\bibinfo{author}{{Stephan}, A.~P.} \emph{et~al.}
\newblock \bibinfo{journal}{\bibinfo{title}{{The Fate of Binaries in the
  Galactic Center: The Mundane and the Exotic}}}.
\newblock {\emph{\JournalTitle{The Astrophysical Journal}}}
  \textbf{\bibinfo{volume}{878}}, \bibinfo{pages}{58},
  \doiprefix\url{10.3847/1538-4357/ab1e4d} (\bibinfo{year}{2019}).
\newblock \eprint{1903.00010}.

\bibitem{2019ApJ...881...20R}
\bibinfo{author}{{Rasskazov}, A.} \& \bibinfo{author}{{Kocsis}, B.}
\newblock \bibinfo{journal}{\bibinfo{title}{{The Rate of Stellar Mass Black
  Hole Scattering in Galactic Nuclei}}}.
\newblock {\emph{\JournalTitle{The Astrophysical Journal}}}
  \textbf{\bibinfo{volume}{881}}, \bibinfo{pages}{20},
  \doiprefix\url{10.3847/1538-4357/ab2c74} (\bibinfo{year}{2019}).
\newblock \eprint{1902.03242}.

\bibitem{2020ApJ...891...47A}
\bibinfo{author}{{Arca Sedda}, M.}
\newblock \bibinfo{journal}{\bibinfo{title}{{Birth, Life, and Death of Black
  Hole Binaries around Supermassive Black Holes: Dynamical Evolution of
  Gravitational Wave Sources}}}.
\newblock {\emph{\JournalTitle{The Astrophysical Journal}}}
  \textbf{\bibinfo{volume}{891}}, \bibinfo{pages}{47},
  \doiprefix\url{10.3847/1538-4357/ab723b} (\bibinfo{year}{2020}).
\newblock \eprint{2002.04037}.

\bibitem{2021MNRAS.505..339M}
\bibinfo{author}{{Mapelli}, M.} \emph{et~al.}
\newblock \bibinfo{journal}{\bibinfo{title}{{Hierarchical black hole mergers in
  young, globular and nuclear star clusters: the effect of metallicity, spin
  and cluster properties}}}.
\newblock {\emph{\JournalTitle{Monthly Notices of the Royal Astronomical
  Society}}} \textbf{\bibinfo{volume}{505}}, \bibinfo{pages}{339--358},
  \doiprefix\url{10.1093/mnras/stab1334} (\bibinfo{year}{2021}).
\newblock \eprint{2103.05016}.

\bibitem{2021ApJ...917...76W}
\bibinfo{author}{{Wang}, H.}, \bibinfo{author}{{Stephan}, A.~P.},
  \bibinfo{author}{{Naoz}, S.}, \bibinfo{author}{{Hoang}, B.-M.} \&
  \bibinfo{author}{{Breivik}, K.}
\newblock \bibinfo{journal}{\bibinfo{title}{{Gravitational-wave Signatures from
  Compact Object Binaries in the Galactic Center}}}.
\newblock {\emph{\JournalTitle{The Astrophysical Journal}}}
  \textbf{\bibinfo{volume}{917}}, \bibinfo{pages}{76},
  \doiprefix\url{10.3847/1538-4357/ac088d} (\bibinfo{year}{2021}).
\newblock \eprint{2010.15841}.

\bibitem{2015ApJ...806..263D}
\bibinfo{author}{{Dominik}, M.} \emph{et~al.}
\newblock \bibinfo{journal}{\bibinfo{title}{{Double Compact Objects III:
  Gravitational-wave Detection Rates}}}.
\newblock {\emph{\JournalTitle{The Astrophysical Journal}}}
  \textbf{\bibinfo{volume}{806}}, \bibinfo{pages}{263},
  \doiprefix\url{10.1088/0004-637X/806/2/263} (\bibinfo{year}{2015}).
\newblock \eprint{1405.7016}.

\bibitem{2016MNRAS.458.2634M}
\bibinfo{author}{{Mandel}, I.} \& \bibinfo{author}{{de Mink}, S.~E.}
\newblock \bibinfo{journal}{\bibinfo{title}{{Merging binary black holes formed
  through chemically homogeneous evolution in short-period stellar binaries}}}.
\newblock {\emph{\JournalTitle{Monthly Notices of the Royal Astronomical
  Society}}} \textbf{\bibinfo{volume}{458}}, \bibinfo{pages}{2634--2647},
  \doiprefix\url{10.1093/mnras/stw379} (\bibinfo{year}{2016}).
\newblock \eprint{1601.00007}.

\bibitem{2016A&A...588A..50M}
\bibinfo{author}{{Marchant}, P.}, \bibinfo{author}{{Langer}, N.},
  \bibinfo{author}{{Podsiadlowski}, P.}, \bibinfo{author}{{Tauris}, T.~M.} \&
  \bibinfo{author}{{Moriya}, T.~J.}
\newblock \bibinfo{journal}{\bibinfo{title}{{A new route towards merging
  massive black holes}}}.
\newblock {\emph{\JournalTitle{Astronomy and Astrophysics}}}
  \textbf{\bibinfo{volume}{588}}, \bibinfo{pages}{A50},
  \doiprefix\url{10.1051/0004-6361/201628133} (\bibinfo{year}{2016}).
\newblock \eprint{1601.03718}.

\bibitem{2017ApJ...841...77A}
\bibinfo{author}{{Antonini}, F.}, \bibinfo{author}{{Toonen}, S.} \&
  \bibinfo{author}{{Hamers}, A.~S.}
\newblock \bibinfo{journal}{\bibinfo{title}{{Binary Black Hole Mergers from
  Field Triples: Properties, Rates, and the Impact of Stellar Evolution}}}.
\newblock {\emph{\JournalTitle{The Astrophysical Journal}}}
  \textbf{\bibinfo{volume}{841}}, \bibinfo{pages}{77},
  \doiprefix\url{10.3847/1538-4357/aa6f5e} (\bibinfo{year}{2017}).
\newblock \eprint{1703.06614}.

\bibitem{2017ApJ...836...39S}
\bibinfo{author}{{Silsbee}, K.} \& \bibinfo{author}{{Tremaine}, S.}
\newblock \bibinfo{journal}{\bibinfo{title}{{Lidov-Kozai Cycles with
  Gravitational Radiation: Merging Black Holes in Isolated Triple Systems}}}.
\newblock {\emph{\JournalTitle{The Astrophysical Journal}}}
  \textbf{\bibinfo{volume}{836}}, \bibinfo{pages}{39},
  \doiprefix\url{10.3847/1538-4357/aa5729} (\bibinfo{year}{2017}).
\newblock \eprint{1608.07642}.

\bibitem{2018MNRAS.481.1908K}
\bibinfo{author}{{Kruckow}, M.~U.}, \bibinfo{author}{{Tauris}, T.~M.},
  \bibinfo{author}{{Langer}, N.}, \bibinfo{author}{{Kramer}, M.} \&
  \bibinfo{author}{{Izzard}, R.~G.}
\newblock \bibinfo{journal}{\bibinfo{title}{{Progenitors of gravitational wave
  mergers: binary evolution with the stellar grid-based code COMBINE}}}.
\newblock {\emph{\JournalTitle{Monthly Notices of the Royal Astronomical
  Society}}} \textbf{\bibinfo{volume}{481}}, \bibinfo{pages}{1908--1949},
  \doiprefix\url{10.1093/mnras/sty2190} (\bibinfo{year}{2018}).
\newblock \eprint{1801.05433}.

\bibitem{2018MNRAS.480.2011G}
\bibinfo{author}{{Giacobbo}, N.} \& \bibinfo{author}{{Mapelli}, M.}
\newblock \bibinfo{journal}{\bibinfo{title}{{The progenitors of compact-object
  binaries: impact of metallicity, common envelope and natal kicks}}}.
\newblock {\emph{\JournalTitle{Monthly Notices of the Royal Astronomical
  Society}}} \textbf{\bibinfo{volume}{480}}, \bibinfo{pages}{2011--2030},
  \doiprefix\url{10.1093/mnras/sty1999} (\bibinfo{year}{2018}).
\newblock \eprint{1806.00001}.

\bibitem{2019MNRAS.490.3740N}
\bibinfo{author}{{Neijssel}, C.~J.} \emph{et~al.}
\newblock \bibinfo{journal}{\bibinfo{title}{{The effect of the
  metallicity-specific star formation history on double compact object
  mergers}}}.
\newblock {\emph{\JournalTitle{Monthly Notices of the Royal Astronomical
  Society}}} \textbf{\bibinfo{volume}{490}}, \bibinfo{pages}{3740--3759},
  \doiprefix\url{10.1093/mnras/stz2840} (\bibinfo{year}{2019}).
\newblock \eprint{1906.08136}.

\bibitem{2019MNRAS.485..889S}
\bibinfo{author}{{Spera}, M.} \emph{et~al.}
\newblock \bibinfo{journal}{\bibinfo{title}{{Merging black hole binaries with
  the SEVN code}}}.
\newblock {\emph{\JournalTitle{Monthly Notices of the Royal Astronomical
  Society}}} \textbf{\bibinfo{volume}{485}}, \bibinfo{pages}{889--907},
  \doiprefix\url{10.1093/mnras/stz359} (\bibinfo{year}{2019}).
\newblock \eprint{1809.04605}.

\bibitem{2010ApJ...718.1400F}
\bibinfo{author}{{Finn}, L.~S.} \& \bibinfo{author}{{Lommen}, A.~N.}
\newblock \bibinfo{journal}{\bibinfo{title}{{Detection, Localization, and
  Characterization of Gravitational Wave Bursts in a Pulsar Timing Array}}}.
\newblock {\emph{\JournalTitle{The Astrophysical Journal}}}
  \textbf{\bibinfo{volume}{718}}, \bibinfo{pages}{1400--1415},
  \doiprefix\url{10.1088/0004-637X/718/2/1400} (\bibinfo{year}{2010}).
\newblock \eprint{1004.3499}.

\bibitem{PhysRevD.80.104009}
\bibinfo{author}{Cutler, C.} \& \bibinfo{author}{Holz, D.~E.}
\newblock \bibinfo{journal}{\bibinfo{title}{Ultrahigh precision cosmology from
  gravitational waves}}.
\newblock {\emph{\JournalTitle{Phys. Rev. D}}} \textbf{\bibinfo{volume}{80}},
  \bibinfo{pages}{104009}, \doiprefix\url{10.1103/PhysRevD.80.104009}
  (\bibinfo{year}{2009}).

\bibitem{2017arXiv170200786A}
\bibinfo{author}{{Amaro-Seoane et al.}}
\newblock \bibinfo{journal}{\bibinfo{title}{{Laser Interferometer Space
  Antenna}}}.
\newblock {\emph{\JournalTitle{arXiv e-prints}}}
  \bibinfo{pages}{arXiv:1702.00786} (\bibinfo{year}{2017}).
\newblock \eprint{1702.00786}.

\bibitem{2022LRR....25....1M}
\bibinfo{author}{{Mandel}, I.} \& \bibinfo{author}{{Broekgaarden}, F.~S.}
\newblock \bibinfo{journal}{\bibinfo{title}{{Rates of compact object
  coalescences}}}.
\newblock {\emph{\JournalTitle{Living Reviews in Relativity}}}
  \textbf{\bibinfo{volume}{25}}, \bibinfo{pages}{1},
  \doiprefix\url{10.1007/s41114-021-00034-3} (\bibinfo{year}{2022}).
\newblock \eprint{2107.14239}.

\bibitem{2021ApJ...910..152Z}
\bibinfo{author}{{Zevin}, M.} \emph{et~al.}
\newblock \bibinfo{journal}{\bibinfo{title}{{One Channel to Rule Them All?
  Constraining the Origins of Binary Black Holes Using Multiple Formation
  Pathways}}}.
\newblock {\emph{\JournalTitle{The Astrophysical Journal}}}
  \textbf{\bibinfo{volume}{910}}, \bibinfo{pages}{152},
  \doiprefix\url{10.3847/1538-4357/abe40e} (\bibinfo{year}{2021}).
\newblock \eprint{2011.10057}.

\bibitem{2023PhRvX..13a1048A}
\bibinfo{author}{{Abbott}, R.} \emph{et~al.}
\newblock \bibinfo{journal}{\bibinfo{title}{{Population of Merging Compact
  Binaries Inferred Using Gravitational Waves through GWTC-3}}}.
\newblock {\emph{\JournalTitle{Physical Review X}}}
  \textbf{\bibinfo{volume}{13}}, \bibinfo{pages}{011048},
  \doiprefix\url{10.1103/PhysRevX.13.011048} (\bibinfo{year}{2023}).
\newblock \eprint{2111.03634}.

\bibitem{2018CQGra..35e4004M}
\bibinfo{author}{{Mandel}, I.}, \bibinfo{author}{{Sesana}, A.} \&
  \bibinfo{author}{{Vecchio}, A.}
\newblock \bibinfo{journal}{\bibinfo{title}{{The astrophysical science case for
  a decihertz gravitational-wave detector}}}.
\newblock {\emph{\JournalTitle{Classical and Quantum Gravity}}}
  \textbf{\bibinfo{volume}{35}}, \bibinfo{pages}{054004},
  \doiprefix\url{10.1088/1361-6382/aaa7e0} (\bibinfo{year}{2018}).
\newblock \eprint{1710.11187}.

\bibitem{2016ApJ...819L..17B}
\bibinfo{author}{{Bellovary}, J.~M.}, \bibinfo{author}{{Mac Low}, M.-M.},
  \bibinfo{author}{{McKernan}, B.} \& \bibinfo{author}{{Ford}, K.~E.~S.}
\newblock \bibinfo{journal}{\bibinfo{title}{{Migration Traps in Disks around
  Supermassive Black Holes}}}.
\newblock {\emph{\JournalTitle{The Astrophysical Journal Letters}}}
  \textbf{\bibinfo{volume}{819}}, \bibinfo{pages}{L17},
  \doiprefix\url{10.3847/2041-8205/819/2/L17} (\bibinfo{year}{2016}).
\newblock \eprint{1511.00005}.

\bibitem{2022PhRvD.106f4056S}
\bibinfo{author}{{Sberna}, L.} \emph{et~al.}
\newblock \bibinfo{journal}{\bibinfo{title}{{Observing GW190521-like binary
  black holes and their environment with LISA}}}.
\newblock {\emph{\JournalTitle{Physical Review D}}}
  \textbf{\bibinfo{volume}{106}}, \bibinfo{pages}{064056},
  \doiprefix\url{10.1103/PhysRevD.106.064056} (\bibinfo{year}{2022}).
\newblock \eprint{2205.08550}.

\bibitem{2020PhRvD.101f3002T}
\bibinfo{author}{{Tamanini}, N.}, \bibinfo{author}{{Klein}, A.},
  \bibinfo{author}{{Bonvin}, C.}, \bibinfo{author}{{Barausse}, E.} \&
  \bibinfo{author}{{Caprini}, C.}
\newblock \bibinfo{journal}{\bibinfo{title}{{Peculiar acceleration of
  stellar-origin black hole binaries: Measurement and biases with LISA}}}.
\newblock {\emph{\JournalTitle{Physical Review D}}}
  \textbf{\bibinfo{volume}{101}}, \bibinfo{pages}{063002},
  \doiprefix\url{10.1103/PhysRevD.101.063002} (\bibinfo{year}{2020}).
\newblock \eprint{1907.02018}.

\bibitem{1992PhRvD..46.5236F}
\bibinfo{author}{{Finn}, L.~S.}
\newblock \bibinfo{journal}{\bibinfo{title}{{Detection, measurement, and
  gravitational radiation}}}.
\newblock {\emph{\JournalTitle{Physical Review D}}}
  \textbf{\bibinfo{volume}{46}}, \bibinfo{pages}{5236--5249},
  \doiprefix\url{10.1103/PhysRevD.46.5236} (\bibinfo{year}{1992}).
\newblock \eprint{gr-qc/9209010}.

\bibitem{1994PhRvD..49.2658C}
\bibinfo{author}{{Cutler}, C.} \& \bibinfo{author}{{Flanagan}, {\'E}.~E.}
\newblock \bibinfo{journal}{\bibinfo{title}{{Gravitational waves from merging
  compact binaries: How accurately can one extract the binary's parameters from
  the inspiral waveform?}}}
\newblock {\emph{\JournalTitle{Physical Review D}}}
  \textbf{\bibinfo{volume}{49}}, \bibinfo{pages}{2658--2697},
  \doiprefix\url{10.1103/PhysRevD.49.2658} (\bibinfo{year}{1994}).
\newblock \eprint{gr-qc/9402014}.

\bibitem{2008PhRvD..77d2001V}
\bibinfo{author}{{Vallisneri}, M.}
\newblock \bibinfo{journal}{\bibinfo{title}{{Use and abuse of the Fisher
  information matrix in the assessment of gravitational-wave
  parameter-estimation prospects}}}.
\newblock {\emph{\JournalTitle{Physical Review D}}}
  \textbf{\bibinfo{volume}{77}}, \bibinfo{pages}{042001},
  \doiprefix\url{10.1103/PhysRevD.77.042001} (\bibinfo{year}{2008}).
\newblock \eprint{gr-qc/0703086}.

\bibitem{emcee}
\bibinfo{author}{{Foreman-Mackey}, D.}, \bibinfo{author}{{Hogg}, D.~W.},
  \bibinfo{author}{{Lang}, D.} \& \bibinfo{author}{{Goodman}, J.}
\newblock \bibinfo{journal}{\bibinfo{title}{{emcee: The MCMC Hammer}}}.
\newblock {\emph{\JournalTitle{Publications of the Astronomical Society of the
  Pacific}}} \textbf{\bibinfo{volume}{125}}, \bibinfo{pages}{306},
  \doiprefix\url{10.1086/670067} (\bibinfo{year}{2013}).
\newblock \eprint{1202.3665}.

\bibitem{Goodman2010}
\bibinfo{author}{{Goodman}, J.} \& \bibinfo{author}{{Weare}, J.}
\newblock \bibinfo{journal}{\bibinfo{title}{{Ensemble samplers with affine
  invariance}}}.
\newblock {\emph{\JournalTitle{Communications in Applied Mathematics and
  Computational Science}}} \textbf{\bibinfo{volume}{5}},
  \bibinfo{pages}{65--80}, \doiprefix\url{10.2140/camcos.2010.5.65}
  (\bibinfo{year}{2010}).

\bibitem{corner}
\bibinfo{author}{Foreman-Mackey, D.}
\newblock \bibinfo{journal}{\bibinfo{title}{corner.py: Scatterplot matrices in
  python}}.
\newblock {\emph{\JournalTitle{The Journal of Open Source Software}}}
  \textbf{\bibinfo{volume}{1}}, \bibinfo{pages}{24},
  \doiprefix\url{10.21105/joss.00024} (\bibinfo{year}{2016}).

\bibitem{2014blanchet}
\bibinfo{author}{{Blanchet}, L.}
\newblock \bibinfo{journal}{\bibinfo{title}{{Gravitational Radiation from
  Post-Newtonian Sources and Inspiralling Compact Binaries}}}.
\newblock {\emph{\JournalTitle{Living Reviews in Relativity}}}
  \textbf{\bibinfo{volume}{17}}, \bibinfo{pages}{2},
  \doiprefix\url{10.12942/lrr-2014-2} (\bibinfo{year}{2014}).
\newblock \eprint{1310.1528}.

\bibitem{MaggioreMichele2018GWV2}
\bibinfo{author}{Maggiore, M.}
\newblock \emph{\bibinfo{title}{{Gravitational Waves: Volume 2: Astrophysics
  and Cosmology}}} (\bibinfo{publisher}{Oxford University Press},
  \bibinfo{year}{2018}).

\bibitem{1979ApJ...234.1100D}
\bibinfo{author}{{Detweiler}, S.}
\newblock \bibinfo{journal}{\bibinfo{title}{{Pulsar timing measurements and the
  search for gravitational waves}}}.
\newblock {\emph{\JournalTitle{The Astrophysical Journal}}}
  \textbf{\bibinfo{volume}{234}}, \bibinfo{pages}{1100--1104},
  \doiprefix\url{10.1086/157593} (\bibinfo{year}{1979}).

\bibitem{1978SvA....22...36S}
\bibinfo{author}{{Sazhin}, M.~V.}
\newblock \bibinfo{journal}{\bibinfo{title}{{Opportunities for detecting
  ultralong gravitational waves}}}.
\newblock {\emph{\JournalTitle{Sov. Astron.}}} \textbf{\bibinfo{volume}{22}},
  \bibinfo{pages}{36--38} (\bibinfo{year}{1978}).

\bibitem{1982MNRAS.199..659M}
\bibinfo{author}{{Mashhoon}, B.}
\newblock \bibinfo{journal}{\bibinfo{title}{{On the contribution of a
  stochastic background of gravitationnal radiation to the timing noise of
  pulsars.}}}
\newblock {\emph{\JournalTitle{Monthly Notices of the Royal Astronomical
  Society}}} \textbf{\bibinfo{volume}{199}}, \bibinfo{pages}{659--666},
  \doiprefix\url{10.1093/mnras/199.3.659} (\bibinfo{year}{1982}).

\bibitem{1983MNRAS.203..945B}
\bibinfo{author}{{Bertotti}, B.}, \bibinfo{author}{{Carr}, B.~J.} \&
  \bibinfo{author}{{Rees}, M.~J.}
\newblock \bibinfo{journal}{\bibinfo{title}{{Limits from the timing of pulsars
  on the cosmic gravitational wave background.}}}
\newblock {\emph{\JournalTitle{Monthly Notices of the Royal Astronomical
  Society}}} \textbf{\bibinfo{volume}{203}}, \bibinfo{pages}{945--954},
  \doiprefix\url{10.1093/mnras/203.4.945} (\bibinfo{year}{1983}).

\bibitem{1983ApJ...265L..39H}
\bibinfo{author}{{Hellings}, R.~W.} \& \bibinfo{author}{{Downs}, G.~S.}
\newblock \bibinfo{journal}{\bibinfo{title}{{Upper limits on the isotropic
  gravitational radiation background from pulsar timing analysis.}}}
\newblock {\emph{\JournalTitle{The Astrophysical Journal Letters}}}
  \textbf{\bibinfo{volume}{265}}, \bibinfo{pages}{L39--L42},
  \doiprefix\url{10.1086/183954} (\bibinfo{year}{1983}).

\bibitem{1990ApJ...361..300F}
\bibinfo{author}{{Foster}, R.~S.} \& \bibinfo{author}{{Backer}, D.~C.}
\newblock \bibinfo{journal}{\bibinfo{title}{{Constructing a Pulsar Timing
  Array}}}.
\newblock {\emph{\JournalTitle{The Astrophysical Journal}}}
  \textbf{\bibinfo{volume}{361}}, \bibinfo{pages}{300},
  \doiprefix\url{10.1086/169195} (\bibinfo{year}{1990}).

\bibitem{1994ApJ...428..713K}
\bibinfo{author}{{Kaspi}, V.~M.}, \bibinfo{author}{{Taylor}, J.~H.} \&
  \bibinfo{author}{{Ryba}, M.~F.}
\newblock \bibinfo{journal}{\bibinfo{title}{{High-Precision Timing of
  Millisecond Pulsars. III. Long-Term Monitoring of PSRs B1855+09 and
  B1937+21}}}.
\newblock {\emph{\JournalTitle{The Astrophysical Journal}}}
  \textbf{\bibinfo{volume}{428}}, \bibinfo{pages}{713},
  \doiprefix\url{10.1086/174280} (\bibinfo{year}{1994}).

\bibitem{2005ApJ...625L.123J}
\bibinfo{author}{{Jenet}, F.~A.}, \bibinfo{author}{{Hobbs}, G.~B.},
  \bibinfo{author}{{Lee}, K.~J.} \& \bibinfo{author}{{Manchester}, R.~N.}
\newblock \bibinfo{journal}{\bibinfo{title}{{Detecting the Stochastic
  Gravitational Wave Background Using Pulsar Timing}}}.
\newblock {\emph{\JournalTitle{The Astrophysical Journal Letters}}}
  \textbf{\bibinfo{volume}{625}}, \bibinfo{pages}{L123--L126},
  \doiprefix\url{10.1086/431220} (\bibinfo{year}{2005}).
\newblock \eprint{astro-ph/0504458}.

\bibitem{2006ApJ...653.1571J}
\bibinfo{author}{{Jenet}, F.~A.} \emph{et~al.}
\newblock \bibinfo{journal}{\bibinfo{title}{{Upper Bounds on the Low-Frequency
  Stochastic Gravitational Wave Background from Pulsar Timing Observations:
  Current Limits and Future Prospects}}}.
\newblock {\emph{\JournalTitle{The Astrophysical Journal}}}
  \textbf{\bibinfo{volume}{653}}, \bibinfo{pages}{1571--1576},
  \doiprefix\url{10.1086/508702} (\bibinfo{year}{2006}).
\newblock \eprint{astro-ph/0609013}.

\bibitem{2009MNRAS.394.1945H}
\bibinfo{author}{{Hobbs}, G.} \emph{et~al.}
\newblock \bibinfo{journal}{\bibinfo{title}{{TEMPO2: a new pulsar timing
  package - III. Gravitational wave simulation}}}.
\newblock {\emph{\JournalTitle{Monthly Notices of the Royal Astronomical
  Society}}} \textbf{\bibinfo{volume}{394}}, \bibinfo{pages}{1945--1955},
  \doiprefix\url{10.1111/j.1365-2966.2009.14391.x} (\bibinfo{year}{2009}).
\newblock \eprint{0901.0592}.

\bibitem{1963PhRv..131..435P}
\bibinfo{author}{{Peters}, P.~C.} \& \bibinfo{author}{{Mathews}, J.}
\newblock \bibinfo{journal}{\bibinfo{title}{{Gravitational Radiation from Point
  Masses in a Keplerian Orbit}}}.
\newblock {\emph{\JournalTitle{Physical Review}}}
  \textbf{\bibinfo{volume}{131}}, \bibinfo{pages}{435--440},
  \doiprefix\url{10.1103/PhysRev.131.435} (\bibinfo{year}{1963}).

\bibitem{MaggioreMichele2008GWV1}
\bibinfo{author}{Maggiore, M.}
\newblock \emph{\bibinfo{title}{{Gravitational Waves: Volume 1: Theory and
  Experiments}}} (\bibinfo{publisher}{Oxford University Press},
  \bibinfo{year}{2007}).

\bibitem{2019PhRvX...9c1040A}
\bibinfo{author}{{Abbott}, B.~P.} \emph{et~al.}
\newblock \bibinfo{journal}{\bibinfo{title}{{GWTC-1: A Gravitational-Wave
  Transient Catalog of Compact Binary Mergers Observed by LIGO and Virgo during
  the First and Second Observing Runs}}}.
\newblock {\emph{\JournalTitle{Physical Review X}}}
  \textbf{\bibinfo{volume}{9}}, \bibinfo{pages}{031040},
  \doiprefix\url{10.1103/PhysRevX.9.031040} (\bibinfo{year}{2019}).
\newblock \eprint{1811.12907}.

\bibitem{2019ApJ...882L..24A}
\bibinfo{author}{{Abbott}, B.~P.} \emph{et~al.}
\newblock \bibinfo{journal}{\bibinfo{title}{{Binary Black Hole Population
  Properties Inferred from the First and Second Observing Runs of Advanced LIGO
  and Advanced Virgo}}}.
\newblock {\emph{\JournalTitle{The Astrophysical Journal Letters}}}
  \textbf{\bibinfo{volume}{882}}, \bibinfo{pages}{L24},
  \doiprefix\url{10.3847/2041-8213/ab3800} (\bibinfo{year}{2019}).
\newblock \eprint{1811.12940}.

\bibitem{2021PhRvX..11b1053A}
\bibinfo{author}{{Abbott}, R.} \emph{et~al.}
\newblock \bibinfo{journal}{\bibinfo{title}{{GWTC-2: Compact Binary
  Coalescences Observed by LIGO and Virgo during the First Half of the Third
  Observing Run}}}.
\newblock {\emph{\JournalTitle{Physical Review X}}}
  \textbf{\bibinfo{volume}{11}}, \bibinfo{pages}{021053},
  \doiprefix\url{10.1103/PhysRevX.11.021053} (\bibinfo{year}{2021}).
\newblock \eprint{2010.14527}.

\bibitem{2021arXiv211103606T}
\bibinfo{author}{{The LIGO Scientific Collaboration}} \emph{et~al.}
\newblock \bibinfo{journal}{\bibinfo{title}{{GWTC-3: Compact Binary
  Coalescences Observed by LIGO and Virgo During the Second Part of the Third
  Observing Run}}}.
\newblock {\emph{\JournalTitle{arXiv e-prints}}}
  \bibinfo{pages}{arXiv:2111.03606}, \doiprefix\url{10.48550/arXiv.2111.03606}
  (\bibinfo{year}{2021}).
\newblock \eprint{2111.03606}.

\bibitem{doi:10.1146/annurev-astro-081811-125615}
\bibinfo{author}{Madau, P.} \& \bibinfo{author}{Dickinson, M.}
\newblock \bibinfo{journal}{\bibinfo{title}{Cosmic star-formation history}}.
\newblock {\emph{\JournalTitle{Annual Review of Astronomy and Astrophysics}}}
  \textbf{\bibinfo{volume}{52}}, \bibinfo{pages}{415--486},
  \doiprefix\url{10.1146/annurev-astro-081811-125615} (\bibinfo{year}{2014}).
\newblock \eprint{https://doi.org/10.1146/annurev-astro-081811-125615}.

\bibitem{2022ApJ...931...17V}
\bibinfo{author}{{van Son}, L.~A.~C.} \emph{et~al.}
\newblock \bibinfo{journal}{\bibinfo{title}{{The Redshift Evolution of the
  Binary Black Hole Merger Rate: A Weighty Matter}}}.
\newblock {\emph{\JournalTitle{The Astrophysical Journal}}}
  \textbf{\bibinfo{volume}{931}}, \bibinfo{pages}{17},
  \doiprefix\url{10.3847/1538-4357/ac64a3} (\bibinfo{year}{2022}).
\newblock \eprint{2110.01634}.

\bibitem{2021arXiv210801045T}
\bibinfo{author}{{The LIGO Scientific Collaboration}} \emph{et~al.}
\newblock \bibinfo{journal}{\bibinfo{title}{{GWTC-2.1: Deep Extended Catalog of
  Compact Binary Coalescences Observed by LIGO and Virgo During the First Half
  of the Third Observing Run}}}.
\newblock {\emph{\JournalTitle{arXiv e-prints}}}
  \bibinfo{pages}{arXiv:2108.01045}, \doiprefix\url{10.48550/arXiv.2108.01045}
  (\bibinfo{year}{2021}).
\newblock \eprint{2108.01045}.

\bibitem{GW150914}
\bibinfo{author}{{Abbott, B.~P. et al.}}
\newblock \bibinfo{journal}{\bibinfo{title}{{Observation of Gravitational Waves
  from a Binary Black Hole Merger}}}.
\newblock {\emph{\JournalTitle{Physical Review Letters}}}
  \textbf{\bibinfo{volume}{116}}, \bibinfo{pages}{061102},
  \doiprefix\url{10.1103/PhysRevLett.116.061102} (\bibinfo{year}{2016}).
\newblock \eprint{1602.03837}.

\bibitem{2011CQGra..28i4011K}
\bibinfo{author}{{Kawamura}, S.} \emph{et~al.}
\newblock \bibinfo{journal}{\bibinfo{title}{{The Japanese space gravitational
  wave antenna: DECIGO}}}.
\newblock {\emph{\JournalTitle{Classical and Quantum Gravity}}}
  \textbf{\bibinfo{volume}{28}}, \bibinfo{pages}{094011},
  \doiprefix\url{10.1088/0264-9381/28/9/094011} (\bibinfo{year}{2011}).

\bibitem{2021PTEP.2021eA105K}
\bibinfo{author}{{Kawamura}, S.} \emph{et~al.}
\newblock \bibinfo{journal}{\bibinfo{title}{{Current status of space
  gravitational wave antenna DECIGO and B-DECIGO}}}.
\newblock {\emph{\JournalTitle{Progress of Theoretical and Experimental
  Physics}}} \textbf{\bibinfo{volume}{2021}}, \bibinfo{pages}{05A105},
  \doiprefix\url{10.1093/ptep/ptab019} (\bibinfo{year}{2021}).
\newblock \eprint{2006.13545}.

\bibitem{PhysRevD.72.083005}
\bibinfo{author}{Crowder, J.} \& \bibinfo{author}{Cornish, N.~J.}
\newblock \bibinfo{journal}{\bibinfo{title}{Beyond lisa: Exploring future
  gravitational wave missions}}.
\newblock {\emph{\JournalTitle{Phys. Rev. D}}} \textbf{\bibinfo{volume}{72}},
  \bibinfo{pages}{083005}, \doiprefix\url{10.1103/PhysRevD.72.083005}
  (\bibinfo{year}{2005}).

\bibitem{2006CQGra..23.4887H}
\bibinfo{author}{{Harry}, G.~M.}, \bibinfo{author}{{Fritschel}, P.},
  \bibinfo{author}{{Shaddock}, D.~A.}, \bibinfo{author}{{Folkner}, W.} \&
  \bibinfo{author}{{Phinney}, E.~S.}
\newblock \bibinfo{journal}{\bibinfo{title}{{Laser interferometry for the Big
  Bang Observer}}}.
\newblock {\emph{\JournalTitle{Classical and Quantum Gravity}}}
  \textbf{\bibinfo{volume}{23}}, \bibinfo{pages}{4887--4894},
  \doiprefix\url{10.1088/0264-9381/23/15/008} (\bibinfo{year}{2006}).

\bibitem{2006CQGra..23..C01H}
\bibinfo{author}{{Harry}, G.~M.}, \bibinfo{author}{{Fritschel}, P.},
  \bibinfo{author}{{Shaddock}, D.~A.}, \bibinfo{author}{{Folkner}, W.} \&
  \bibinfo{author}{{Phinney}, E.~S.}
\newblock \bibinfo{journal}{\bibinfo{title}{{CORRIGENDUM: Laser interferometry
  for the Big Bang Observer}}}.
\newblock {\emph{\JournalTitle{Classical and Quantum Gravity}}}
  \textbf{\bibinfo{volume}{23}}, \bibinfo{pages}{C01},
  \doiprefix\url{10.1088/0264-9381/23/24/C01} (\bibinfo{year}{2006}).

\bibitem{2023LRR....26....2A}
\bibinfo{author}{{Amaro-Seoane et al.}}
\newblock \bibinfo{journal}{\bibinfo{title}{{Astrophysics with the Laser
  Interferometer Space Antenna}}}.
\newblock {\emph{\JournalTitle{Living Reviews in Relativity}}}
  \textbf{\bibinfo{volume}{26}}, \bibinfo{pages}{2},
  \doiprefix\url{10.1007/s41114-022-00041-y} (\bibinfo{year}{2023}).
\newblock \eprint{2203.06016}.

\bibitem{2019CQGra..36j5011R}
\bibinfo{author}{{Robson}, T.}, \bibinfo{author}{{Cornish}, N.~J.} \&
  \bibinfo{author}{{Liu}, C.}
\newblock \bibinfo{journal}{\bibinfo{title}{{The construction and use of LISA
  sensitivity curves}}}.
\newblock {\emph{\JournalTitle{Classical and Quantum Gravity}}}
  \textbf{\bibinfo{volume}{36}}, \bibinfo{pages}{105011},
  \doiprefix\url{10.1088/1361-6382/ab1101} (\bibinfo{year}{2019}).
\newblock \eprint{1803.01944}.

\bibitem{2019PhRvD.100d3003W}
\bibinfo{author}{{Wang}, H.-T.} \emph{et~al.}
\newblock \bibinfo{journal}{\bibinfo{title}{{Science with the TianQin
  observatory: Preliminary results on massive black hole binaries}}}.
\newblock {\emph{\JournalTitle{Physical Review D}}}
  \textbf{\bibinfo{volume}{100}}, \bibinfo{pages}{043003},
  \doiprefix\url{10.1103/PhysRevD.100.043003} (\bibinfo{year}{2019}).
\newblock \eprint{1902.04423}.

\bibitem{2013CQGra..30v4010M}
\bibinfo{author}{{Manchester}, R.~N.} \& \bibinfo{author}{{IPTA}}.
\newblock \bibinfo{journal}{\bibinfo{title}{{The International Pulsar Timing
  Array}}}.
\newblock {\emph{\JournalTitle{Classical and Quantum Gravity}}}
  \textbf{\bibinfo{volume}{30}}, \bibinfo{pages}{224010},
  \doiprefix\url{10.1088/0264-9381/30/22/224010} (\bibinfo{year}{2013}).
\newblock \eprint{1309.7392}.

\bibitem{2015MNRAS.453.2576L}
\bibinfo{author}{{Lentati}, L.} \emph{et~al.}
\newblock \bibinfo{journal}{\bibinfo{title}{{European Pulsar Timing Array
  limits on an isotropic stochastic gravitational-wave background}}}.
\newblock {\emph{\JournalTitle{Monthly Notices of the Royal Astronomical
  Society}}} \textbf{\bibinfo{volume}{453}}, \bibinfo{pages}{2576--2598},
  \doiprefix\url{10.1093/mnras/stv1538} (\bibinfo{year}{2015}).
\newblock \eprint{1504.03692}.

\bibitem{2015aska.confE..37J}
\bibinfo{author}{{Janssen}, G.} \emph{et~al.}
\newblock \bibinfo{title}{{Gravitational Wave Astronomy with the SKA}}.
\newblock In \emph{\bibinfo{booktitle}{Advancing Astrophysics with the Square
  Kilometre Array (AASKA14)}}, \bibinfo{pages}{37},
  \doiprefix\url{10.22323/1.215.0037} (\bibinfo{year}{2015}).
\newblock \eprint{1501.00127}.

\bibitem{2020ApJ...905L..34A}
\bibinfo{author}{{Arzoumanian}, Z.} \emph{et~al.}
\newblock \bibinfo{journal}{\bibinfo{title}{{The NANOGrav 12.5 yr Data Set:
  Search for an Isotropic Stochastic Gravitational-wave Background}}}.
\newblock {\emph{\JournalTitle{The Astrophysical Journal Letters}}}
  \textbf{\bibinfo{volume}{905}}, \bibinfo{pages}{L34},
  \doiprefix\url{10.3847/2041-8213/abd401} (\bibinfo{year}{2020}).
\newblock \eprint{2009.04496}.

\bibitem{2023arXiv230103608A}
\bibinfo{author}{{Arzoumanian}, Z.} \emph{et~al.}
\newblock \bibinfo{journal}{\bibinfo{title}{{The NANOGrav 12.5-year Data Set:
  Bayesian Limits on Gravitational Waves from Individual Supermassive Black
  Hole Binaries}}}.
\newblock {\emph{\JournalTitle{arXiv e-prints}}}
  \bibinfo{pages}{arXiv:2301.03608}, \doiprefix\url{10.48550/arXiv.2301.03608}
  (\bibinfo{year}{2023}).
\newblock \eprint{2301.03608}.

\bibitem{2013ApJ...777...44J}
\bibinfo{author}{{Ju}, W.}, \bibinfo{author}{{Greene}, J.~E.},
  \bibinfo{author}{{Rafikov}, R.~R.}, \bibinfo{author}{{Bickerton}, S.~J.} \&
  \bibinfo{author}{{Badenes}, C.}
\newblock \bibinfo{journal}{\bibinfo{title}{{Search for Supermassive Black Hole
  Binaries in the Sloan Digital Sky Survey Spectroscopic Sample}}}.
\newblock {\emph{\JournalTitle{The Astrophysical Journal}}}
  \textbf{\bibinfo{volume}{777}}, \bibinfo{pages}{44},
  \doiprefix\url{10.1088/0004-637X/777/1/44} (\bibinfo{year}{2013}).
\newblock \eprint{1306.4987}.

\bibitem{2022LRR....25....3B}
\bibinfo{author}{{Bogdanovi{\'c}}, T.}, \bibinfo{author}{{Miller}, M.~C.} \&
  \bibinfo{author}{{Blecha}, L.}
\newblock \bibinfo{journal}{\bibinfo{title}{{Electromagnetic counterparts to
  massive black-hole mergers}}}.
\newblock {\emph{\JournalTitle{Living Reviews in Relativity}}}
  \textbf{\bibinfo{volume}{25}}, \bibinfo{pages}{3},
  \doiprefix\url{10.1007/s41114-022-00037-8} (\bibinfo{year}{2022}).
\newblock \eprint{2109.03262}.

\bibitem{2022MNRAS.511.6143Z}
\bibinfo{author}{{Zwick}, L.}, \bibinfo{author}{{Derdzinski}, A.},
  \bibinfo{author}{{Garg}, M.}, \bibinfo{author}{{Capelo}, P.~R.} \&
  \bibinfo{author}{{Mayer}, L.}
\newblock \bibinfo{journal}{\bibinfo{title}{{Dirty waveforms: multiband
  harmonic content of gas-embedded gravitational wave sources}}}.
\newblock {\emph{\JournalTitle{Monthly Notices of the Royal Astronomical
  Society}}} \textbf{\bibinfo{volume}{511}}, \bibinfo{pages}{6143--6159},
  \doiprefix\url{10.1093/mnras/stac299} (\bibinfo{year}{2022}).
\newblock \eprint{2110.09097}.

\bibitem{2010MNRAS.406.2267F}
\bibinfo{author}{{Fakhouri}, O.}, \bibinfo{author}{{Ma}, C.-P.} \&
  \bibinfo{author}{{Boylan-Kolchin}, M.}
\newblock \bibinfo{journal}{\bibinfo{title}{{The merger rates and mass assembly
  histories of dark matter haloes in the two Millennium simulations}}}.
\newblock {\emph{\JournalTitle{Monthly Notices of the Royal Astronomical
  Society}}} \textbf{\bibinfo{volume}{406}}, \bibinfo{pages}{2267--2278},
  \doiprefix\url{10.1111/j.1365-2966.2010.16859.x} (\bibinfo{year}{2010}).
\newblock \eprint{1001.2304}.

\bibitem{2013MNRAS.428..421S}
\bibinfo{author}{{Shankar}, F.}, \bibinfo{author}{{Weinberg}, D.~H.} \&
  \bibinfo{author}{{Miralda-Escud{\'e}}, J.}
\newblock \bibinfo{journal}{\bibinfo{title}{{Accretion-driven evolution of
  black holes: Eddington ratios, duty cycles and active galaxy fractions}}}.
\newblock {\emph{\JournalTitle{Monthly Notices of the Royal Astronomical
  Society}}} \textbf{\bibinfo{volume}{428}}, \bibinfo{pages}{421--446},
  \doiprefix\url{10.1093/mnras/sts026} (\bibinfo{year}{2013}).
\newblock \eprint{1111.3574}.

\bibitem{2009MNRAS.394.1109C}
\bibinfo{author}{{Croton}, D.~J.}
\newblock \bibinfo{journal}{\bibinfo{title}{{A simple model to link the
  properties of quasars to the properties of dark matter haloes out to high
  redshift}}}.
\newblock {\emph{\JournalTitle{Monthly Notices of the Royal Astronomical
  Society}}} \textbf{\bibinfo{volume}{394}}, \bibinfo{pages}{1109--1119},
  \doiprefix\url{10.1111/j.1365-2966.2009.14429.x} (\bibinfo{year}{2009}).
\newblock \eprint{0901.4104}.

\bibitem{2021MNRAS.507.4274M}
\bibinfo{author}{{Marasco}, A.} \emph{et~al.}
\newblock \bibinfo{journal}{\bibinfo{title}{{A universal relation between the
  properties of supermassive black holes, galaxies, and dark matter haloes}}}.
\newblock {\emph{\JournalTitle{Monthly Notices of the Royal Astronomical
  Society}}} \textbf{\bibinfo{volume}{507}}, \bibinfo{pages}{4274--4293},
  \doiprefix\url{10.1093/mnras/stab2317} (\bibinfo{year}{2021}).
\newblock \eprint{2105.10508}.

\bibitem{2023MNRAS.tmp.1567B}
\bibinfo{author}{{Bansal}, A.}, \bibinfo{author}{{Ichiki}, K.},
  \bibinfo{author}{{Tashiro}, H.} \& \bibinfo{author}{{Matsuoka}, Y.}
\newblock \bibinfo{journal}{\bibinfo{title}{{Evolution of the mass relation
  between supermassive black holes and dark matter halos across the cosmic
  time}}}.
\newblock {\emph{\JournalTitle{Monthly Notices of the Royal Astronomical
  Society}}} \doiprefix\url{10.1093/mnras/stad1608} (\bibinfo{year}{2023}).
\newblock \eprint{2206.01443}.

\bibitem{2018ricarte}
\bibinfo{author}{{Ricarte}, A.} \& \bibinfo{author}{{Natarajan}, P.}
\newblock \bibinfo{journal}{\bibinfo{title}{{The observational signatures of
  supermassive black hole seeds}}}.
\newblock {\emph{\JournalTitle{Monthly Notices of the Royal Astronomical
  Society}}} \textbf{\bibinfo{volume}{481}}, \bibinfo{pages}{3278--3292},
  \doiprefix\url{10.1093/mnras/sty2448} (\bibinfo{year}{2018}).
\newblock \eprint{1809.01177}.

\bibitem{2019MNRAS.487..275G}
\bibinfo{author}{{Georgakakis}, A.} \emph{et~al.}
\newblock \bibinfo{journal}{\bibinfo{title}{{Exploring the halo occupation of
  AGN using dark-matter cosmological simulations}}}.
\newblock {\emph{\JournalTitle{Monthly Notices of the Royal Astronomical
  Society}}} \textbf{\bibinfo{volume}{487}}, \bibinfo{pages}{275--295},
  \doiprefix\url{10.1093/mnras/sty3454} (\bibinfo{year}{2019}).
\newblock \eprint{1812.04025}.

\bibitem{2015ApJ...810...49V}
\bibinfo{author}{{Vasiliev}, E.}, \bibinfo{author}{{Antonini}, F.} \&
  \bibinfo{author}{{Merritt}, D.}
\newblock \bibinfo{journal}{\bibinfo{title}{{The Final-parsec Problem in the
  Collisionless Limit}}}.
\newblock {\emph{\JournalTitle{The Astrophysical Journal}}}
  \textbf{\bibinfo{volume}{810}}, \bibinfo{pages}{49},
  \doiprefix\url{10.1088/0004-637X/810/1/49} (\bibinfo{year}{2015}).
\newblock \eprint{1505.05480}.

\bibitem{2017ApJ...840...31D}
\bibinfo{author}{{Dosopoulou}, F.} \& \bibinfo{author}{{Antonini}, F.}
\newblock \bibinfo{journal}{\bibinfo{title}{{Dynamical Friction and the
  Evolution of Supermassive Black Hole Binaries: The Final Hundred-parsec
  Problem}}}.
\newblock {\emph{\JournalTitle{The Astrophysical Journal}}}
  \textbf{\bibinfo{volume}{840}}, \bibinfo{pages}{31},
  \doiprefix\url{10.3847/1538-4357/aa6b58} (\bibinfo{year}{2017}).
\newblock \eprint{1611.06573}.

\bibitem{2016PhRvD..93b4003K}
\bibinfo{author}{{Klein}, A.} \emph{et~al.}
\newblock \bibinfo{journal}{\bibinfo{title}{{Science with the space-based
  interferometer eLISA: Supermassive black hole binaries}}}.
\newblock {\emph{\JournalTitle{Physical Review D}}}
  \textbf{\bibinfo{volume}{93}}, \bibinfo{pages}{024003},
  \doiprefix\url{10.1103/PhysRevD.93.024003} (\bibinfo{year}{2016}).
\newblock \eprint{1511.05581}.

\bibitem{2024arXiv240207571C}
\bibinfo{author}{{Colpi}, M.} \emph{et~al.}
\newblock \bibinfo{journal}{\bibinfo{title}{{LISA Definition Study Report}}}.
\newblock {\emph{\JournalTitle{arXiv e-prints}}}
  \bibinfo{pages}{arXiv:2402.07571}, \doiprefix\url{10.48550/arXiv.2402.07571}
  (\bibinfo{year}{2024}).
\newblock \eprint{2402.07571}.

\bibitem{2018MNRAS.477.1822M}
\bibinfo{author}{{Moster}, B.~P.}, \bibinfo{author}{{Naab}, T.} \&
  \bibinfo{author}{{White}, S. D.~M.}
\newblock \bibinfo{journal}{\bibinfo{title}{{EMERGE - an empirical model for
  the formation of galaxies since z {\ensuremath{\sim}} 10}}}.
\newblock {\emph{\JournalTitle{Monthly Notices of the Royal Astronomical
  Society}}} \textbf{\bibinfo{volume}{477}}, \bibinfo{pages}{1822--1852},
  \doiprefix\url{10.1093/mnras/sty655} (\bibinfo{year}{2018}).
\newblock \eprint{1705.05373}.

\bibitem{2020A&A...634A.135G}
\bibinfo{author}{{Girelli}, G.} \emph{et~al.}
\newblock \bibinfo{journal}{\bibinfo{title}{{The stellar-to-halo mass relation
  over the past 12 Gyr. I. Standard {\ensuremath{\Lambda}}CDM model}}}.
\newblock {\emph{\JournalTitle{Astronomy and Astrophysics}}}
  \textbf{\bibinfo{volume}{634}}, \bibinfo{pages}{A135},
  \doiprefix\url{10.1051/0004-6361/201936329} (\bibinfo{year}{2020}).
\newblock \eprint{2001.02230}.

\end{thebibliography}

\end{document}